\documentclass[superscriptaddress,twocolumn,showpacs,preprintnumbers,amsmath,amssymb,aps]{revtex4-1}
\usepackage{epsfig}
\pdfoutput=1

\usepackage{subfig}
\usepackage[T1]{fontenc}
\usepackage[latin9]{inputenc}
\usepackage{amsmath,amssymb,amsthm,amsfonts}
\usepackage{ mathrsfs }
\usepackage{slashed}
\usepackage{graphicx}
\usepackage{color,rotating}
\usepackage{dsfont}
\usepackage{setspace}
\usepackage{verbatim}
\usepackage{fancyhdr}
\usepackage{hyperref}
\usepackage{mathtools}

\def\Fig#1{Fig.~\ref{#1}}

\def\s0#1#2{\mbox{\small{$ \frac{#1}{#2} $}}}
\def\0#1#2{\frac{#1}{#2}}

\usepackage{caption}
\newcommand{\sumint}{\int\hspace{-4.8mm}\sum}

\newcommand{\beq}{\begin{equation}}
\newcommand{\eeq}{\end{equation}}
\newcommand{\beqa}{\begin{eqnarray}}
\newcommand{\eeqa}{\end{eqnarray}}

\graphicspath{{./figures/}}

\newcommand{\bea}{\begin{eqnarray}}
\newcommand{\eea}{\end{eqnarray}}
\newcommand{\Eq}[1]{Eq.~(\ref{#1})}

\definecolor{darkgreen}{rgb}{0,0.6,0}
\definecolor{gray}{rgb}{.7,.7,.7}

\def\eq#1{(\ref{#1})}
\def\Eq#1{Eq.~(\ref{#1})}
\newcommand {\apgt} {\ {\raise-.5ex\hbox{$\buildrel>\over\sim$}}\ }
\newcommand {\aplt} {\ {\raise-.5ex\hbox{$\buildrel<\over\sim$}}\ }

\def\s0#1#2{\mbox{\small{$ \frac{#1}{#2} $}}}
\def\0#1#2{\frac{#1}{#2}}

\def\Co{{ c}}

\newcommand{\Tr}{\mathrm{Tr}}

\newcommand{\be}{\begin{eqnarray}}
\newcommand{\ee}{\end{eqnarray}}
\newcommand{\dm}{{\rm d}}

\newcommand{\rmi}{\mathrm{i}}

\newcommand{\feyn}[1]{
  \setbox0=\hbox{\ensuremath{#1}}
  \hbox to\wd0{\hbox to0pt{\hbox to\wd0{\hss/\hss}\hss}\box0}}

\DeclareMathOperator*{\SumInt}{%
\mathchoice%
  {\ooalign{$\displaystyle\sum$\cr\hidewidth$\displaystyle\int$\hidewidth\cr}}
  {\ooalign{\raisebox{.14\height}{\scalebox{.7}{$\textstyle\sum$}}\cr\hidewidth$\textstyle\int$\hidewidth\cr}}
  {\ooalign{\raisebox{.2\height}{\scalebox{.6}{$\scriptstyle\sum$}}\cr$\scriptstyle\int$\cr}}
  {\ooalign{\raisebox{.2\height}{\scalebox{.6}{$\scriptstyle\sum$}}\cr$\scriptstyle\int$\cr}}
}

\begin{document}
\title{Magnetic catalysis and inverse magnetic catalysis in QCD} \vspace{1.5 true cm} 

\author{Niklas Mueller} \affiliation{Institut f\"ur Theoretische
  Physik, Universit\"at Heidelberg, Philosophenweg 16, 69120
  Heidelberg, Germany}

\author{Jan M. Pawlowski} \affiliation{Institut f\"ur Theoretische
  Physik, Universit\"at Heidelberg, Philosophenweg 16, 69120
  Heidelberg, Germany} \affiliation{ExtreMe Matter Institute EMMI, GSI
  Helmholtzzentrum f\"ur Schwerionenforschung mbH, Planckstr. 1,
  D-64291 Darmstadt, Germany}

\begin{abstract}
  We investigate the effects of strong magnetic fields on the QCD
  phase structure at vanishing density by solving the gluon and quark
  gap equations, and by studying the dynamics of the quark scattering
  with the four-fermi coupling. The chiral crossover temperature as
  well as the chiral condensate are computed. For asymptotically large
  magnetic fields we find magnetic catalysis, while we find inverse
  magnetic catalysis for intermediate magnetic fields. Moreover, for
  large magnetic fields the chiral phase transition for massless
  quarks turns into a crossover.

  The underlying mechanisms are then investigated analytically within
  a few simplifications of the full numerical analysis. We find that a
  combination of gluon screening effects and the weakening of the
  strong coupling is responsible for the phenomenon of inverse
  catalysis. In turn, the magnetic catalysis
  at large magnetic field is already indicated by simple arguments
  based on dimensionality. 
\end{abstract}




\maketitle

\section{Introduction}\noindent%
In recent years there has been a growing interest in the QCD phase
structure in the presence of strong magnetic fields, see e.g.\
\cite{Shovkovy:2012zn,D'Elia:2012tr,Fukushima:2012vr,Kharzeev:2007jp,%
  Fukushima:2008xe,Mueller:2014tea,Andersen:2014xxa,Miransky:2015ava}. Such fields may
play an important role for the physics of the early universe, in
compact stars, and in non-central heavy ion collisions
\cite{Kharzeev:2007jp,Andersen:2014xxa,Basar:2012bp,Basar:2014swa}. 

Despite the rich phenomenology, theoretical predictions are
challenging. Starting from QED, e.g.\
\cite{Gusynin:1995nb,Gusynin:1998zq,Gusynin:1999pq,Gusynin:1995gt} the
influence of magnetic fields onto QCD was investigated in both model
calculations, e.g.\
\cite{Inagaki:2003yi,Boomsma:2009yk,D'Elia:2010nq,Mizher:2010zb,%
 Gatto:2010qs,Gatto:2010pt,Chatterjee:2011ry,Gatto:2012sp,Frasca:2011zn,%
  Ferreira:2014kpa,Ferrer:2014qka,Farias:2014eca,Ayala:2014iba,Ayala:2014gwa,Yu:2014xoa}, such as quark-meson,
Nambu-Jona-Lasinio models and AdS/QCD, e.g.\
\cite{Bergman:2008sg,Johnson:2008vna,Filev:2009xp,Preis:2010cq,%
  Preis:2012fh,Ballon-Bayona:2013cta,Mamo:2015dea}, with functional
renormalisation group methods, e.g.\
\cite{Skokov:2011ib,Fukushima:2012xw,Kamikado:2013pya,Kamikado:2014bua,
  Andersen:2012bq,Andersen:2013swa,Braun:2014fua}, Dyson-Schwinger
equations, e.g.\
\cite{Kojo:2012js,Kojo:2013uua,Watson:2013ghq,Mueller:2014tea}
and in lattice calculations, e.g.\
\cite{Bali:2011qj,Bali:2012zg,Bali:2013esa,Bruckmann:2013oba,%
  Bali:2014kia,Ilgenfritz:2013ara,Bornyakov:2013eya}.

The importance of magnetic fields for chiral symmetry breaking has
been pointed out in \cite{Gusynin:1995nb}. It has been argued that
chiral symmetry breaking is enhanced due to an effective dimensional
reduction, the \textit{magnetic catalysis}. This effect has been 
linked to an increase of the chiral condensate as well as that of the
critical temperature $T_c$ in model studies. Recent lattice results,
\cite{Bali:2011qj,Bali:2012zg,Bali:2013esa,Bornyakov:2013eya}, have shown that while the
chiral condensate indeed is increased, the critical temperature is
decreasing with an increasing magnetic field, at least for small
enough magnetic field strength. This effect has been called
\textit{inverse magnetic catalysis} or \textit{magnetic inhibition},
\cite{Fukushima:2012kc}.

Continuum studies have mainly been performed in low energy fermionic
models, such as the (Polyakov loop enhanced) quark-meson-- or
NJL--model. Hence the reason for the discrepancy has to relate to the
full dynamics of QCD, and in particular the back-reaction of the matter
sector to the gluonic fluctuations. There have been a number of
improvements to these model studies to include QCD dynamics
\cite{Fraga:2013ova,Ferreira:2014kpa,Farias:2014eca,Ayala:2014iba,Ayala:2014gwa,Andersen:2014oaa}. Input
parameters of low energy effective models, such as the four-fermi
coupling, should be determined from the QCD dynamics at larger
scales. At these scales they are sensitive to sufficiently large
external parameters such as temperature, density, or magnetic
fields. This has been emphasized and used in functional
renormalisation group (FRG) studies, see
\cite{Pawlowski:2010ht,Haas:2013qwp,Herbst:2013ufa,Pawlowski:2014aha}.
The dependence of the four-fermi coupling on temperature and magnetic
field effects including gluon screening has been investigated in the
recent FRG-work \cite{Braun:2014fua} of QCD in strong magnetic fields,
where inverse magnetic catalysis at small magnetic fields and a {\it
  delayed magnetic catalysis} at large fields was found, see also
\cite{Mamo:2015dea} for an AdS/QCD computation. 

In the present work we investigate (inverse) magnetic catalysis by
solving the coupled quark and gluon gap equations within the
Dyson-Schwinger (DSE) approach to QCD, and within a FRG study of the
four-fermi coupling based on QCD flows and low energy effective
models. We find magnetic catalysis at large magnetic fields, while
inverse magnetic catalysis takes place at small magnetic fields.

The present work is organized as follows: The gap equations for quark
and gluon propagators at finite temperature and magnetic field in two
flavor QCD are discussed in Section~\ref{sec:gap}.  We discuss the
dependence of the chiral transition temperature $T_c$ on the magnetic
field as well as the magnetic field dependence of the chiral
condensate. In Section~\ref{sec:analytic} the mechanisms behind the
phenomena of magnetic and inverse magnetic catalysis are evaluated
within analytically accessible approximations to the gap equations as
well as to the dynamics of the four-fermi coupling. In this set-up we
are also able to reproduce the lattice results at $eB<1\text{
  GeV}^2$. In summary this provides a complete picture of chiral
symmetry breaking in the presence of magnetic fields in QCD.

\section{Chiral symmetry breaking in large magnetic
  fields}\label{sec:gap} 

We investigate chiral symmetry breaking in the presence of large
magnetic fields within a functional continuum approach. To this end we calculate the
chiral condensate for the two lightest quark flavors
and obtain the critical temperature $T_c$ at finite magnetic field.
 This is done
by solving the gap equations for the quark and gluon propagator in the
presence of a magnetic field using the Ritus method
\cite{Lee:1997zj,Miransky:2002rp,Leung:2005xz,Ayala:2006sv%
  ,Rojas:2008sg,Ritus:1972ky,Ritus:1978cj}. The computations are
performed in the Landau gauge.

\subsection{Quark and gluon gap equations} 
\begin{figure}[t]
\includegraphics[width=.95\columnwidth]{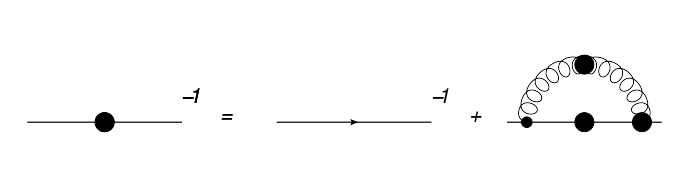}
\caption{Quark Dyson-Schwinger equation. Lines with blobs stand for
  fully dressed propagators, vertices with large blobs stand for fully
  dressed vertices. Lines without blobs stand for classical
  propagators, vertices with small blobs stand for classical
  vertices.}
\label{fig:quarkDSE}
\end{figure}

The gap equation for the quark propagator, see \Fig{fig:quarkDSE},
depends on the gluon propagator and the quark-gluon vertex.  The
former is expanded about the quenched propagator. This expansion has
been successfully used at vanishing temperature, e.g.\
\cite{Fischer:2005en,Nickel:2006kc}, and at finite temperature in
e.g.\ \cite{Braun:2009gm,Fischer:2012vc,Fischer:2013eca,Fischer:2014ata}, 
the reliability of this expansion has been
discussed in \cite{Braun:2014ata}.  The quark-gluon vertex is
estimated with the help of Slavnov-Taylor identities (STIs) from the
quark and gluon propagators. The systematic error of the latter
estimate gives rise to the dominating systematic error, at vanishing
temperature this has been investigated in \cite{Mitter:2014wpa}, a
related upgrade of the vertex will be used in a subsequent work.

\begin{figure}[t]
  \includegraphics[width=.95\columnwidth]{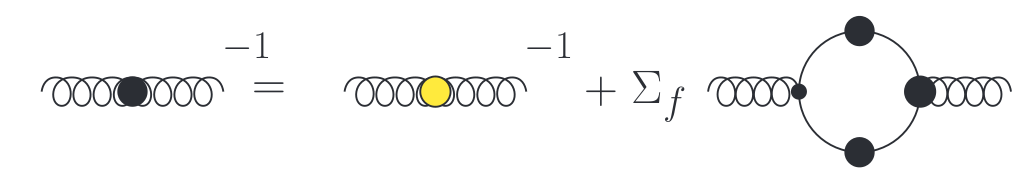}
  \caption{Gluon Dyson-Schwinger equation. The gluon line with the yellow dot 
represents the pure glue loops.}
\label{fig:gluonDSE}
\end{figure}

\noindent%
The inverse quark and gluon propagators, $G_q(q)$ and $G_A(q)$
respectively, read in a tensor decomposition at finite $eB$ and $T$
\begin{align}\nonumber 
  G^{-1}_q(q) &= Z_q(q) \left(
    i\gamma_3q_3+i\gamma_0q_0Z_0+i\gamma_\perp q_\perp
    Z_\perp+M\right) \,, \\[2ex]
  G_A^{-1\,\mu\nu}(q) & = \left(Z^{\ {} }_\parallel \,
    P_\parallel^{\mu\nu} + Z_\perp^{\ {}} \,
    P_\perp^{\mu\nu}\right)q^2 +\01\xi \0{q^\mu q^\nu}{q^2} \,,
\label{eq:Gs}\end{align}
with $P_\parallel^{\mu \nu}=(g_\|^{\mu\nu}-p_\parallel^\mu
p_\parallel^\nu /p_\parallel^2)$ and $P_\perp=P-P_\parallel$, where
$P_{\mu\nu}$ is the transverse projector. The projection
operator $g_\|^{\mu\nu}$ has the property $g_\|^{\mu\nu} p_\|^\mu =
p_\|^\nu$. The Ritus representation \Eq{eq:Gs} for the quark propagator 
is equivalent to the Schwinger proper time method, see e.g. \cite{Chyi:1999fc}.
In the following we will denote $Z_A\equiv Z_\| $ and
concentrate on the Landau gauge, $\xi=0$. The STIs-induced
parametrisation of the quark-gluon vertex is introduced as
\begin{align}
\Gamma^{\mu}(q,p)=\gamma^{\mu}z_{\bar{q}Aq}^\text{DSE}(q,p)\,,
\end{align}
with $z_{\bar{q}Aq}^\text{DSE}(q,p)$ discussed in
Appendix~\ref{app:propsandvertex}.
\noindent%
The quark gap equation can be written in a
compact notation as
\begin{align}
  G_q^{-1}(p)=G_{q,0}^{-1}(p)+ C_f\SumInt_q (g\gamma^\mu) G_q(q)
  \Gamma^\nu(q,p) G_{A}^{\mu\nu}(q')\,,
  \label{eq:skeleton_quarkDSE}
\end{align}
with $q'=q-p$ and $G_{q,0}$ as the bare propagator. The integration
$\SumInt_q$ stands for an integration over momenta, as well as sums
over Matsubara frequencies and Landau levels. The gluon propagator can
be expanded about its pure glue part,
\begin{align}
  G_A^{-1\,\mu\nu}(p)=G_\text{\rm
    glue}^{-1\,\mu\nu}(p)+\Pi_f^{\mu\nu}(p)\,,\label{eq:gluonDSE}
\end{align}
where we have written the fermionic part of the gluon self energy
explicitly, while the gluon and ghost loop contributions are contained
in $G_\text{\rm glue}$. The corresponding DSE for the gluon propagator
within this expansion is depicted \Fig{fig:gluonDSE}. In the following
we consider the back-reaction of the vacuum polarisation on the pure
glue part as small, and approximate
\begin{align}\label{eq:glueYM} 
  G_\text{\rm glue}^{-1\,\mu\nu}(p) \approx G_\text{\rm
    YM}^{-1\,\mu\nu}(p)\,.
\end{align}
At vanishing temperature this has been shown to hold quantitatively
for momenta $q \gtrsim 4\text{ GeV}$, while for smaller momenta this
approximation still holds qualitatively with an error of less than
20\%, see Fig.~6 in \cite{Braun:2014ata}.  Note that for momenta $q
\gtrsim 4\text{ GeV}$ the dominant effect of the unquenching is the
modification of the scales ($\Lambda_{\rm YM}\to \Lambda_{\rm QCD}$)
and the momentum dependence induced by the different
$\beta$-functions. This is well-captured with the above procedure.  In
turn, at lower momentum scale the non-perturbative mass-gap related to
confinement comes into play. The magnetic field leads to a shift in
the momentum dependence such as that of the running coupling, as well
as (additional) mass-gaps in propagators. For both asymptotic regimes
($e B \to 0$ and $ e B \to \infty$) these effects are well-captured
semi-perturbatively and we expect that the approximation
\eq{eq:glueYM} holds well. For the intermediate regime we rely on the
error estimate at zero temperature of about 20\% deduced from
\cite{Braun:2014ata}.

The fermionic vacuum polarisation part $\Pi_f^{\mu\nu}(P)$ reads 
\begin{equation}
  \Pi_f^{\mu\nu}(p)=\012 \text{tr}\SumInt_q(g\gamma^\mu) G_q(q)
  \Gamma^\nu(q,p) G_q(q')\,,
\end{equation}
where the trace includes a sum over the quark flavors. Details of this
expansion can be found in \cite{Mueller:2014tea}.  Here we proceed in
the lowest Landau level approximation, where we write down the most
general tensor decomposition for gluon and quark
propagators. Projecting onto different tensor compositions, we obtain
a coupled set of equations for the dressing functions of the different
tensor components. In the next section we will comment on the relation
of the Dyson-Schwinger equations to other functional expansions and
discuss the numerical solutions to these equations.

\subsection{Skeleton expansion}\label{sec:skeleton}
\begin{figure}[t]
\includegraphics[width=7cm]{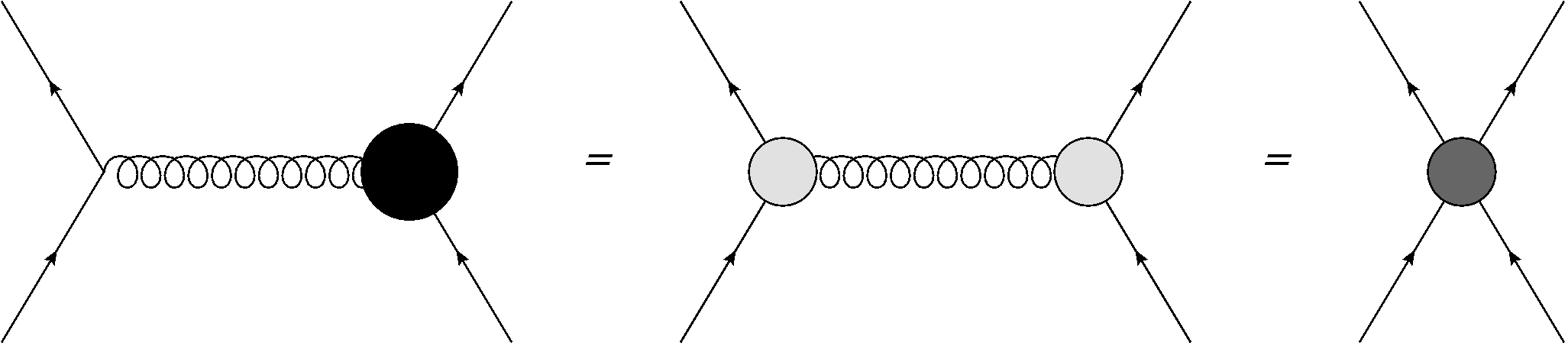}
\caption{Relation of the quark DSE interaction kernel to a 1PI
  skeleton expansion, which in effect induces an effective momentum
  dependent four-fermi vertex.}
\label{fig:DSEkernel}
\end{figure}%
%
%
%
%
%
%
%
Before proceeding to the numerical analysis, we discuss the standard
approximation schemes for the quark-gluon vertex used in the Dyson-Schwinger
framework from a more general point of view. This allows us to
connect the present ans\"atze to the approximations used in gap
equations derived within other functional approaches, such as
functional renormalisation group (FRG) or nPI-approaches. 

DSE studies have made extensively use of the specific input for the
quark-gluon vertex and the YM-gluon propagator in \eq{eq:DSE:vtx} and
\eq{eq:gluonQuenched} and similar truncations with great
success. Since the quark and gluon self energy diagrams, depicted in
\Fig{fig:quarkDSE} and \Fig{fig:gluonDSE}, contain one bare vertex, the correct
renormalisation group behavior and momentum dependence of the
equations must be discussed carefully.  The truncations to the gap
equations \eq{eq:skeleton_quarkDSE} and \eq{eq:gluonDSE} can
actually be very well motivated from a skeleton expansion of the 1PI
effective action, which would yield similar diagrams as in 
\Fig{fig:quarkDSE} and \Fig{fig:gluonDSE}, but with both vertices
dressed. \Fig{fig:DSEkernel} serves to strengthen this motivation as
it becomes clear that all approximations should encode the correct
behavior of the four-fermi interaction, which is at the heart of
chiral symmetry breaking. This allows to consistently reshuffle
functional dependencies in the interaction kernels of the above
equations. 

In turn, the FRG-approach (or nPI effective action) can be used to
systematically derive gap equations in terms of full propagators and
vertices respectively, see e.g.\ \cite{Pawlowski:2005xe}. Here, we
simply note that the 1PI effective action can be written as 
\begin{align}\label{eq:1PI}
  \Gamma[\phi]= \012{ \Tr} \ln \Gamma[\phi] + \int_t \partial_t
  \Gamma_k[\phi]-{\rm terms}\,,
\end{align} 
where $\phi$ encodes all species of fields, the trace in \eq{eq:1PI}
sums over momenta, internal indices and all species of fields
including relative minus signs for fermions ($\psi$ and $\psi^\dagger$
are counted separately), and a logarithmic RG-scale $t=\ln k$. The
RG-scale in \eq{eq:1PI} is an infrared scale. Momenta $p^2 \lesssim
k^2$ are suppressed in $\Gamma_k[\phi]$, and
$\Gamma[\phi]=\Gamma_{k=0}[\phi]$. The second term on the right hand
side of \eq{eq:1PI} is a RG-improvement term which only contains
diagrams with two loops and more in full propagators and vertices. To
see this we discuss the gap equation derived from \eq{eq:1PI}. It
follows by taking the second derivative of \eq{eq:1PI} w.r.t.\ to the
fields. The first term of the right hand side gives the diagrams as in
\Fig{fig:quarkDSE} and \Fig{fig:gluonDSE} with only full vertices (and
additional tadpole diagrams). These diagrams can be iteratively
re-inserted into the RG-improvement term, systematically leading to
higher loop diagrams in full propagators and vertices. Due to its sole
dependence on dressed correlation functions such a diagrammatics 
naturally encodes the momentum- as well as the RG-running on an equal
footing. This also facilitates the consistent renormalisation. Note
however that it comes at the price of an infinite series of loops
diagrams which can be computed systematically. Here we take the
simplest non-trivial approximation which boils down to
\Fig{fig:quarkDSE} and \Fig{fig:gluonDSE} with only full vertices. In
terms of the original gap equation this leads to the relation
\begin{equation}
z_{\bar{q}Aq}^\text{DSE}\approx  \left(z_{\bar{q}Aq}^\text{1PI}\right)^2,
\end{equation}
where $z_{\bar{q}Aq}^\text{1PI}$ is the dressing function of the
1PI-quark gluon vertex. This immediately leads to the standard
DSE-dressing in \eq{eq:DSE:vtx}. Moreover, in our numerical study the
vertices are evaluated at their symmetric momentum point.%

Note that, while the ansatz for $z^\text{DSE}_{\bar{q}gq}$ is indeed
consistent when used in the quark and gluon gap equations, it cannot
be used in functional equations for higher vertices such as the
four-fermi vertex. It is already clear from the discussion above
that a consistent evaluation of renormalisation group running and
momentum dependence must be considered separately for each vertex
equation. 

\subsection{Results}%

We numerically solve the coupled system of quark and gluon functional
equations in the lowest Landau level approximation at finite
temperature. This approximation is valid in the presence of a clear
scale hierarchy with $eB\gg\Lambda_\text{QCD}$. 
We use an ansatz for $\Gamma^\mu$ similar to that used in
Dyson-Schwinger studies, e.g.\ \cite{Mueller:2014tea,Fischer:2010fx},
discussed in appendix \ref{app:propsandvertex}, but adapted for
temperature and magnetic field effects. 

While at large momentum the influence of temperature and magnetic
fields is very small, at large temperatures and magnetic fields the
system is effectively dimensionally reduced and hence the momentum
dependencies corresponding to the absent dimensions vanish. This can be
accounted for if we replace $Q_\perp^2$ by $2|eB|$ once $Q_\perp^2 <
2|eB|$ and $Q_0$ by $2\pi T$ for $Q_0<2\pi T$ as the relevant scale in
the quark gluon vertex, which is consistent with renormalisation group
arguments.  Within this parametrisation we are still left to decide
what exact momentum scale to choose, at which the influence of the external
scales $T$ and $eB$ is small already. We investigate this
question in detail in section \ref{sec:DSEan}.

\begin{figure}[t]
\includegraphics[width=.95\columnwidth]{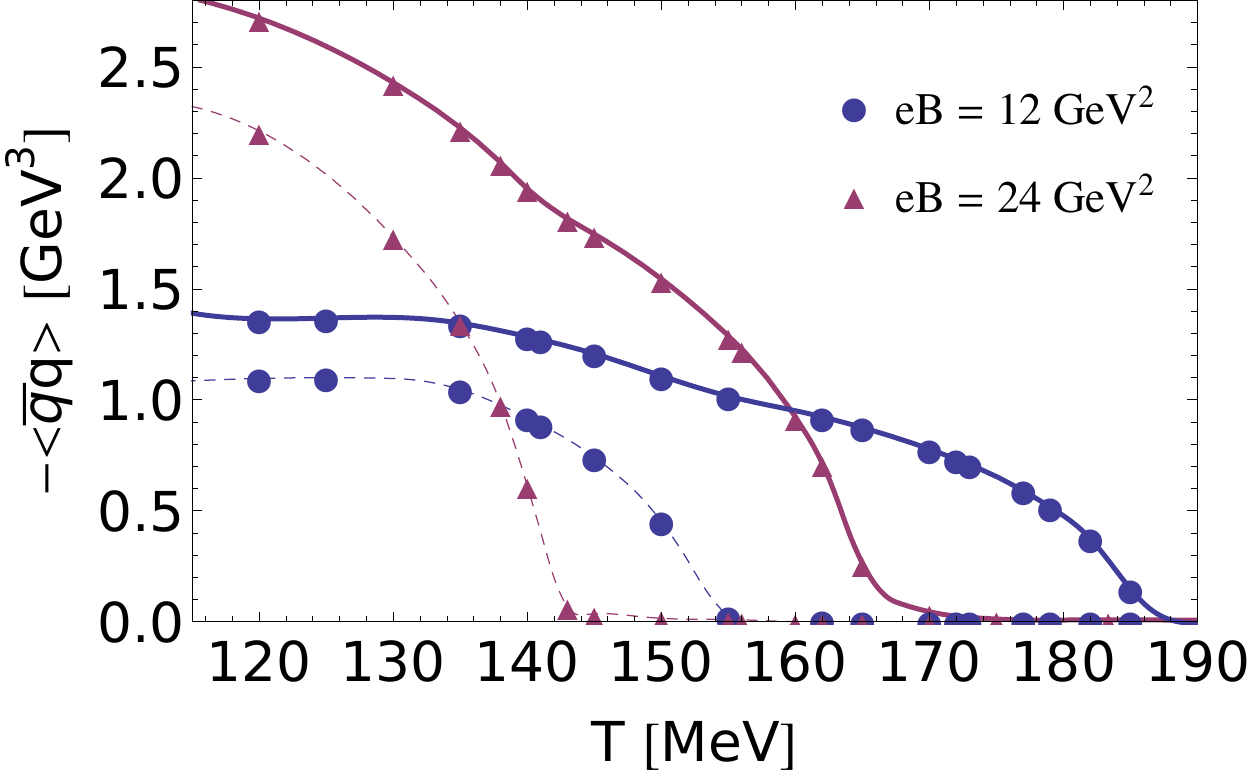}
\caption{Comparison of the chiral condensate (scenario 1) for up
  (continuous lines) and down quark (dashed lines) at $eB=12\text{ GeV}^2$ and $eB=24\text{ GeV}^2$.}
\label{fig:condensate1}
\end{figure}
The gluon propagator deserves some additional attention. It is
decomposed in different polarisation components in the presence of an
external magnetic field, see e.g.\ \cite{Mueller:2014tea}. Apart from
the splitting into longitudinal and transverse components with respect
to the heat bath, there is an additional splitting transverse and
longitudinal to the magnetic field. In the lowest Landau level
approximation only the polarisation subspace projected onto by
$P_\parallel^{\mu \nu}=(g_\parallel^{\mu\nu}-p_\parallel^\mu
p_\parallel^\nu /p_\parallel^2)$ receives contributions from the quark
loop in the self energy, see \cite{Mueller:2014tea}. Note that in
analogy to temperature effects, also the other gluon components must
receive contributions from the interaction with the magnetic field, as
gluon and ghost loops mix different polarisation components. This is
an important difference between QCD and QED. From dimensionality these
contributions are linear in $eB$ at least for asymptotically large
magnetic fields, leaving aside implicit $B$-dependencies via the
vertices. Their full computation is beyond the scope of the present
work.  Here we investigate the following two limiting cases.
\begin{enumerate}
\item \textbf{Scenario 1} We simply neglect the screening effect of
  the magnetic field onto those polarisation components that feel
  magnetic effects only through the Yang-Mills sector in a QED-type
  approximation. This leads to underestimating the effects leading to
  inverse magnetic catalysis and hence an upper limit for $T_c$.
\item \textbf{Scenario 2} For the large magnetic fields discussed
  here, the gluon and ghost loops contributions to the self energy
  must have a similar dependence on $eB$ as the fermionic part. Since
  this sector does not directly contain charged particles, the effect
  of the magnetic field onto the YM-sector is suppressed by powers of
  the involved couplings. Hence, most likely the $B$-dependence is
  much smaller than that from the fermionic sector. As a limiting case we
  will assume the same magnitude of the self energy for all gluon
  components, which is given by the fermionic contributions. With that
  we overestimate the gluon screening effect and obtain a lower limit
  for $T_c$.
\end{enumerate}
Both scenarios give consistent limiting cases for the truncation used
here.

As an order parameter for chiral symmetry breaking we calculate the
chiral condensate as a function of temperature and magnetic field in
two flavor QCD in the limit of vanishing bare quark masses $m_u\approx
m_d\approx 0$. The Ritus method is not reliable for rather small
values of $q_f eB$, with $q_f+2/3$ and $-1/3$ for up and down quark
respectively. We expect the lowest Landau level approximation to be a
good estimate once $eB\gtrsim 4\text{ GeV}^2$ (see
\cite{Mueller:2014tea}) which is also the regime where the
approximation \eq{eq:glueYM} works well for vanishing
temperature. 

\begin{figure}[t]
  \includegraphics[width=.95\columnwidth]{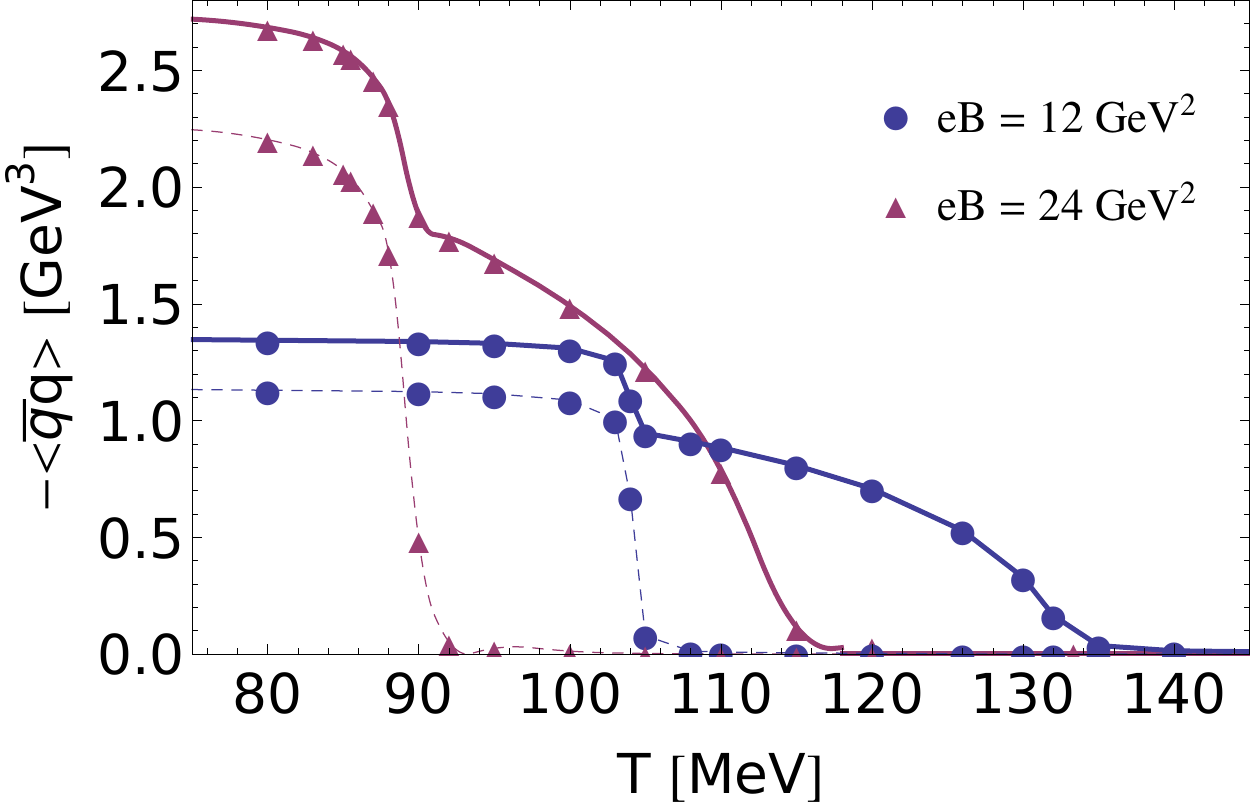}
  \caption{Comparison of the chiral condensate for scenario~2 at
    $eB=12\text{ GeV}^2$ and $eB=24\text{ GeV}^2$.}
\label{fig:condensate2}
\end{figure}

The numerical computation is very demanding in the vicinity of the
phase transition due to the diverging correlation length. This
translates into a numerical error in the critical temperature
indicated by the error bars in the plots. \Fig{fig:condensate1} and
\Fig{fig:condensate2} show the up- and down-quark condensate for
different values of $eB$. The inverse magnetic catalysis effect
  described in \cite{Bali:2013esa,Bali:2012zg} is evident.  While the
chiral condensate still rises with the external field in the low
temperature limit, the transition between chiral broken and symmetric
phase drops. This signals inverse magnetic catalysis as observed on
the lattice, \cite{Bali:2011qj,Bali:2012zg}. Furthermore the phase
transition, which is second order at zero magnetic field turns into a
crossover with growing $eB$, even for vanishing bare quark
masses. This can be understood as magnetic screening: the magnetic
field effectively serves as an infrared cutoff, which inhibits an
infinite correlation length.%
In the present computation in two-flavor QCD, an even more intricate
effect is observed. Up and down quarks come with different electric
charges, therefore the presence of a strong electromagnetic field
breaks isospin explicitly. This results in a non-degenerate chiral
phase transition for the two flavors. Because gluons travel through a
medium filled with both virtual up and down quarks, isospin breaking
effects the self interactions of the quarks, which leads to
interference between the chiral transitions of the two flavors as seen
in \Fig{fig:condensate1} and \Fig{fig:condensate2}. 

This interference can be interpreted as follows. Virtual quark
fluctuations contributing to the gluon screening are suppressed in the
chiral broken phase by the quark mass. Since the down quark undergoes
the chiral phase transition already at lower scales, its fluctuations
are suddenly enhanced due to the vanishing mass in the symmetric
phase. The up quark, while still in its chirally broken phase, is
drastically effected by these enhanced fluctuations, which lead to
reduction of the up quark condensate even below the real phase
transition. 

\begin{figure}[t]
\includegraphics[width=.95\columnwidth]{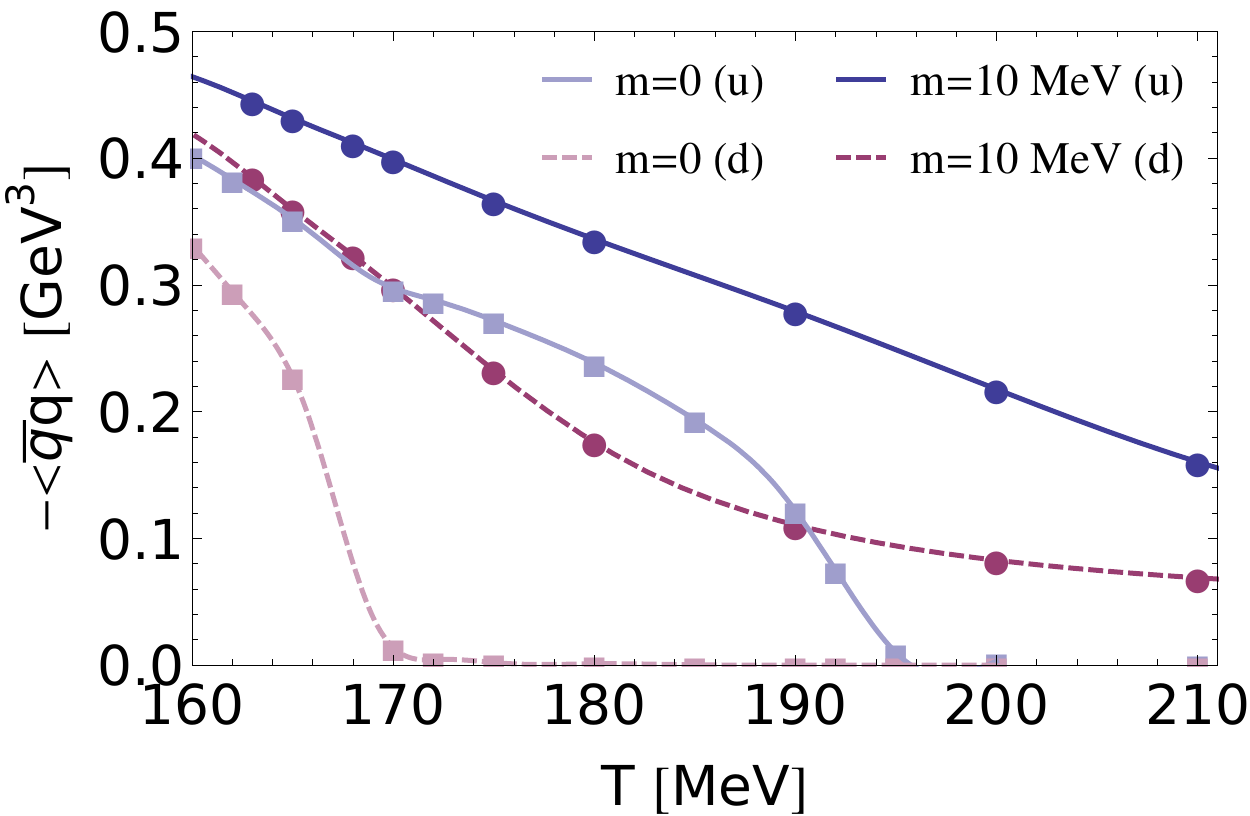}
\caption{Comparison of the chiral condensate at zero bare mass and at
  a finite bare quark mass of $m_u=m_d=10\text{ MeV}$ at $eB=4\text{
    GeV}^2$ in scenario 1.}
\label{fig:Tc_massive}
\end{figure}

It can be seen from \Fig{fig:condensate1} and \Fig{fig:condensate2}
that this effect is more prominent in scenario 2, which should come as
no surprise, as the coupling of the magnetic field to the gauge sector
is probably overestimated here. Nevertheless the isospin induced
chiral transition substructure is observable in the limiting scenario
1 as well, which is a strong indication of its validity. Therefore
this important physical effect might be observable in lattice
calculations, as well. In \cite{Bali:2013esa,Bali:2012zg} the averaged
chiral condensate was investigated at finite quark mass.  However when
we investigate the chiral transition at a bare quark mass of $10
\text{ MeV}$ we find that the interference effect is completely masked
by the crossover behavior as can be seen in \Fig{fig:Tc_massive}. Note
that here the unregularized condensate at finite bare mass is plotted,
hence the offset between the curves.%

\begin{figure}[t]
\includegraphics[width=.95\columnwidth]{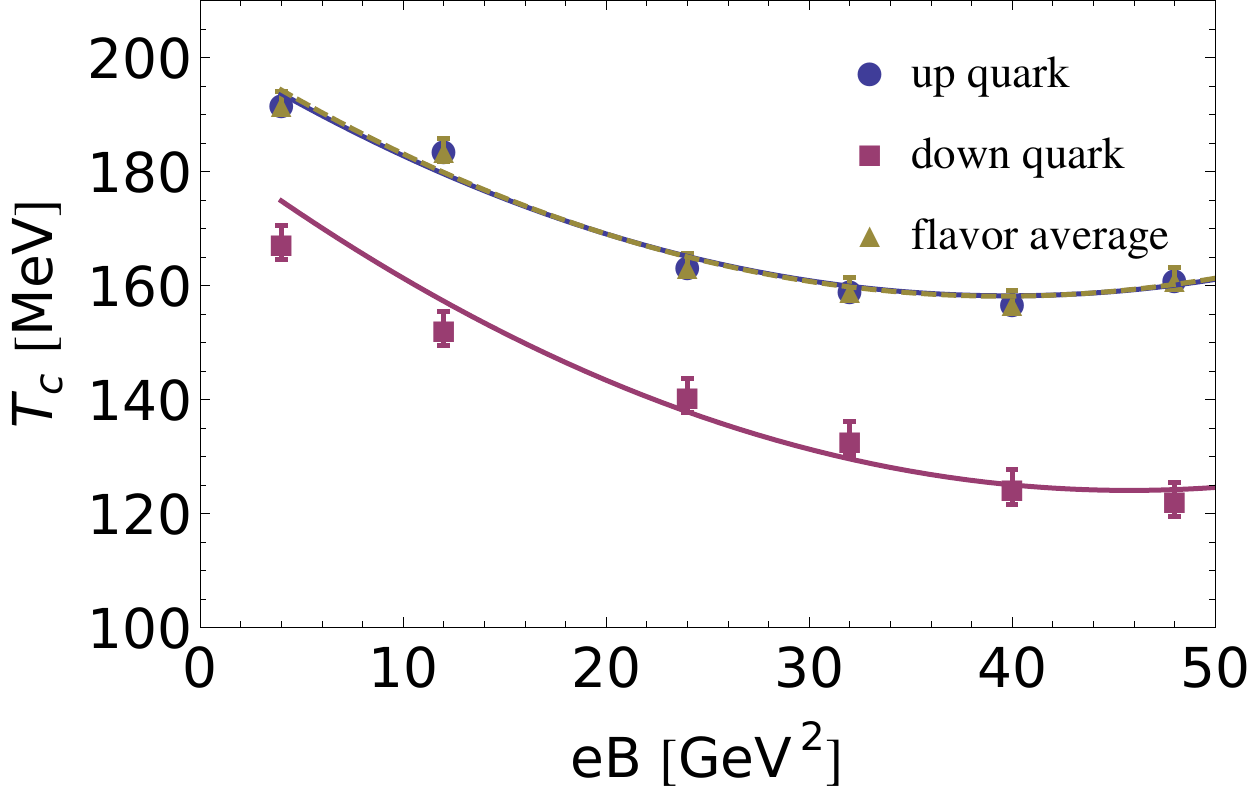}
\caption{Critical temperature obtained from scenario 1 for up quark,
  down quark and from the flavor averaged condensate.}
\label{fig:numericTc_1}
\end{figure}

\begin{figure}[t]
\includegraphics[width=.95\columnwidth]{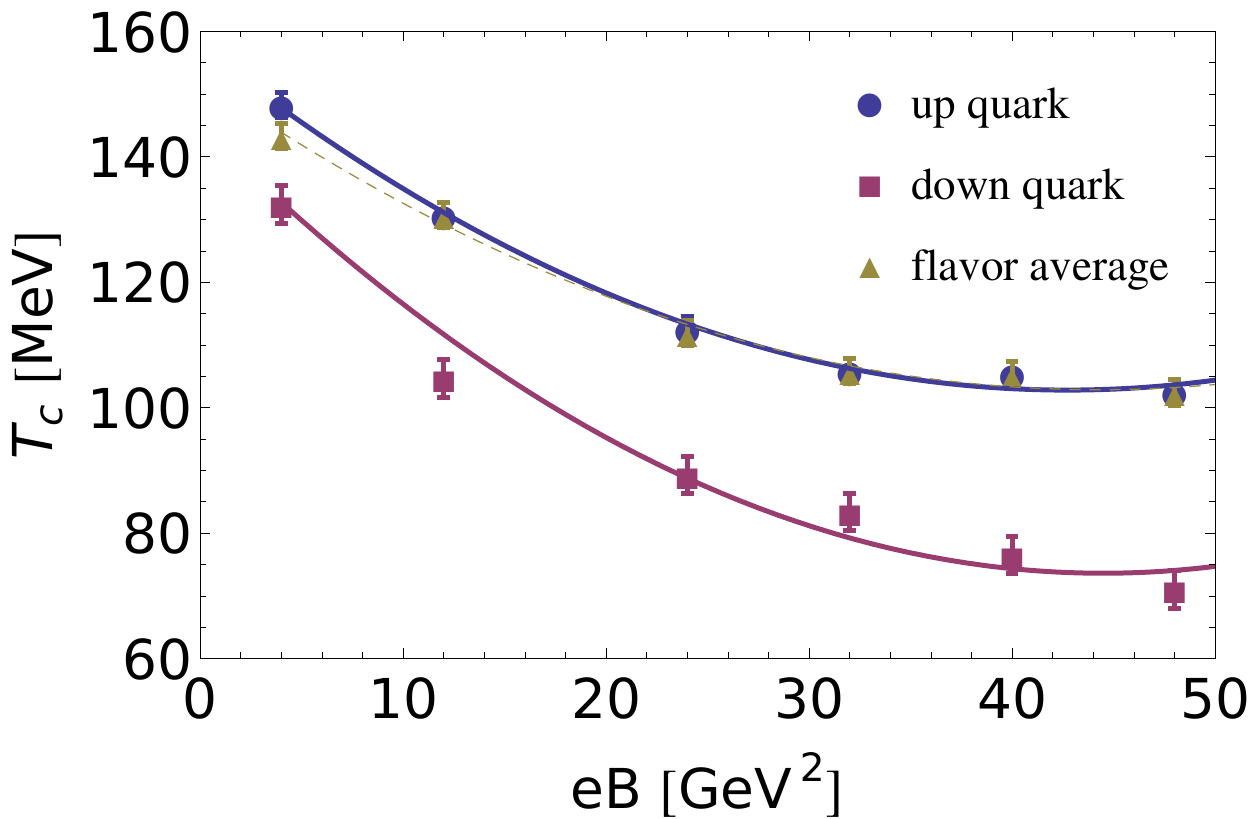}
\caption{Critical temperature obtained from scenario 2.}
\label{fig:numericTc_2}
\end{figure}

In analogy with lattice calculation we define $T_c$ at the inflection
points of the curves shown. In \Fig{fig:numericTc_1} and
\Fig{fig:numericTc_2} the obtained values for $T_c$ for the limiting
cases described by scenario 1 and 2 are shown.
The two curves give lower and upper limits for $T_c$, as discussed
before. The chiral transition temperature is decreasing for a large
range in $eB$ before it seems to saturate for intermediate values in
both scenarios. At very large fields it rises again. 

In accordance with our previous discussions we see that the up and
down quark chiral transitions do not coincide. The transition
temperature from the flavor averaged quark condensate is given in
\Fig{fig:numericTc_1} and \Fig{fig:numericTc_2} as well. As can be
seen from \Fig{fig:condensate1} and \Fig{fig:condensate2} the
transition temperature of the flavor averaged condensate is
essentially determined by the up quark.%

Both scenarios give estimates for the chiral transition temperature,
which differ only quantitatively.  Scenario 1, which underestimates
the magnetic field effects in the gluon sector extrapolates to a
critical temperature at $eB=0$ between $170-210 \text{ MeV}$ with a
turning point between catalysis and inverse catalysis of about
$eB\approx 30\text{ GeV}^2$. On the other hand scenario 2 gives $T_c$
at zero magnetic field of about $140-165\text{ MeV}$ with a turning
point slightly higher than in scenario 1. This is in accordance with
the fact that scenario 2 overestimates the gluonic sector, which is
the source of the inverse catalysis effects. At $B=0$ the chiral phase
transitions for up and down quark coincide. While the continuous lines
in \Fig{fig:numericTc_1} and \Fig{fig:numericTc_2} are obtained from a
fit with a simple quadratic polynomial, reflecting the turnover
behavior at large fields, these should not be mistaken as
extrapolations towards zero. Furthermore the computations have been
performed in the lowest landau level approximation. This leads to an
uncertainty of about 10\% for $B$ smaller than $10\text{ GeV}^2$,
while the qualitative behavior is not effected, as discussed in
\cite{Mueller:2014tea,Miransky:2015ava}.  In the following section we
will see that the behavior of $T_c$ at small $B$ is steeper than just
quadratic. 

It is well known that within approximation schemes such as the one
discussed here, relative fluctuation scales are usually well accounted
for, whereas absolute scales have to be fixed. The position of $T_c$
at $eB=0$ gives us the possibility of identifying absolute scales and
allows to adjust our truncation. We will not be concerned about
matching the exact scale of $T_c$ at zero magnetic field with the
lattice, moreover we will investigate the mechanisms behind the $B-T$
phase structure in greater detail. We will discuss the issue of scales
in the following sections.
%
%
%
%
%
\section{Analytic approaches} \label{sec:analytic}
In the present Section we are specifically interested in the
mechanisms at work in magnetic and inverse magnetic catalysis. To that
end we discuss approximations to the quark gap equation in 
Section~\ref{sec:DSEan}, as well as to the dynamics of the four-fermi coupling or
quark scattering kernel in Section~\ref{sec:QCD-NJL}, that allow for an
analytic approach to chiral symmetry breaking. While the quark gap
equation can be straightforwardly reduced to an analytic form from
that used for the numerical study, the four-fermi coupling is studied
in a renormalisation group approach to QCD, that reduces to an
NJL-type model for low momentum scales.

\subsection{Quark gap equation}\label{sec:DSEan}
The mechanisms behind the phenomena observed in our numerical study
can be analyzed within approximations detailed below, that allow for
an analytic access. These approximations to the gap equation have been
introduced in \cite{Gusynin:1999pq} for QED, and can be extended to
QCD at finite temperature. The self-consistent Dyson-Schwinger
equation for the mass functions reads in lowest Landau level
approximation with zero bare mass
\begin{align}
  M(p_\parallel)=4\pi C_F \sumint\limits_{q_\parallel}\nonumber
  \frac{M(q_\parallel)\Tr{(\Delta(\text{sgn}(eB))\gamma^\mu_\parallel
      \gamma^\nu_\parallel})}{M^2(q_\parallel)+q_\parallel^2}\\[2ex] 
  \int \limits_{k_\perp}\;\alpha_s\exp\left(-\frac{k^2_\perp} {2|eB|}
  \right)\frac{P_{\mu\nu}(k)}{k^2+\Pi(k^2)}\,.
\label{eq:selfgap}\end{align}
Here $\SumInt=T\sum_{n}\int\dm{q_\parallel}/(2\pi)^3$ and
$\Delta(s)=(1+s\sigma^3)/2$. The quark gap equation \eq{eq:selfgap} is
obtained from a skeleton expansion of the effective action, e.g.\
\cite{Fischer:2009tn}, and is nothing but a manifestly renormalisation
group invariant approximation of the above Dyson-Schwinger equations,
see the discussion in Section~\ref{sec:skeleton}. It includes only
dressed vertices. In appendix \ref{app:propsandvertex} we discuss how
the interaction kernels can be related in both pictures. The 1PI
quark-gluon vertex is parametrized as
\begin{align}
  \Gamma_{\bar{q}Aq}^\mu(q^2)=Z_A^{1/2}(q^2)\sqrt{4\pi
    \alpha_s(q^2)}\gamma^\mu_\parallel\,,
\label{eq:vertexskeleton}
\end{align}
The gluon propagator is transversal due to the Landau gauge, and we
allow for a gluonic mass via thermal and magnetic
effects. $M(p_\parallel)$ is a function that is approximately constant
in the IR but falls of rapidly for $p_\parallel^2\ge2|eB|$. Hence, if
we are interested in $M(0)=M_{\rm IR}$ we can write, dividing the
equation by its trivial solution,
\begin{align}\nonumber
 &1-4\pi^2C_FT\sumint\limits_{q_\parallel}^{2eB}\frac{1}{M_{\rm IR}^2+
q_{\parallel,f}^2}\\[2ex] 
 & \hspace{.5cm}\times \int\dm x \frac{\alpha_s\exp{\left(
-x/2|eB|\right)}}{q_{\parallel,b}^2+
    x+\Pi(x,q_{\parallel,b})}\left( 2-\frac{q_{\parallel,b}^2}{q_{\parallel,b}^2
      +x}\right)=0\,.\label{eq:DSEzeros:1}
\end{align}
In \eq{eq:DSEzeros:1} we have introduced $q_{\parallel,b}\equiv (q_3,
2n\pi T)$ and $q_{\parallel,f}\equiv (q_3,2\pi T(n+1/2))$.  Chiral
symmetry breaking is realized once a solution $M_{\rm IR}^2> 0$
exists. Due to the shape of $M(q)$ and the exponential factor
in \eq{eq:DSEzeros:1}, the integrand only has support for $x\lesssim
2|eB|$. In the following we carefully investigate the ingredients
to this self consistent equation and the physical mechanisms, which
are responsible for the intriguing behavior seen in the previous
section.%

Due to the finite support of the integrand, the momenta running
through the vertices are comparable or smaller than the relevant
dimensionful quantities $eB$ and $T^2$. Note that in our numerical
study we have used an ansatz for the quark gluon vertex, that includes
generic $eB$ and $T$ dependencies. Here we utilize the fact that the
running of $\alpha_s$ is dominated by the temperature and magnetic
field scales. We resort to a simple ansatz for
$\alpha_s(Q^2/\Lambda_{\rm QCD}^2)$ based on the analytic coupling
$\alpha_{s,\text{\tiny{HQ}}}$ suggested in \cite{Nesterenko:1999np,Nesterenko:2001st},
see \cite{Christiansen:2014ypa} for an investigation within the
present context. This coupling yields a linear potential such as seen in 
the heavy quark limit.  
\begin{align} 
  \alpha_s(z)=\alpha_{s,\text{\tiny HQ}}(z)\, r_\text{\tiny
    IR}(z)\,,\label{eq:alpha1}
\end{align} 
where
\begin{align} 
  \alpha_{s,\text{\tiny HQ}}(z)=\0{1}{\beta_0}\frac{z^2-1}{z^2\log(z^2)}\,,\label{eq:alphaNP}
\end{align} 
with $\beta_0=(33-2N_f)/12\pi$ and
\begin{align}
  z^2=\0{\lambda_B 2eB+\lambda_T(2\pi T)^2}{\Lambda_{\rm QCD}^2}\,,
  \label{eq:scales}
\end{align}%
with coefficients $\lambda_T$, $\lambda_B$, which are of order one.
These coefficients determine the point at which $eB$ or $T$ dominate
momentum scales. For the relevant magnetic fields and temperatures the
running of the coupling with temperature is very small compared to the
running with $eB$. We use an ansatz for the infrared behavior of the
vertex, which is parametrized in $r_\text{IR}$. Here we use
\begin{align}
  r_\text{\tiny IR}(z^2)=\frac{z^4}{(z^2+b^2)^2}\left(
    1+\frac{c^2}{z^2+b^2}\right)\,,
\label{eq:rIR}
\end{align}
which scales with $\propto z^4$ for $z\rightarrow 0$, and approaches
unity in the perturbative regime. \Eq{eq:alpha1} reproduces the
correct behavior of the full quark gluon vertex in
\eq{eq:vertexskeleton}. We leave $b$ and $c$ as parameters which allow
us to model the infrared behavior of the quark gluon vertex. Our
ansatz for \eq{eq:rIR} is motivated from the quantitative
renormalisation group study of quenched QCD in \cite{Mitter:2014wpa},
which we use to determine $b$ and $c$.  We get
\begin{align}
  b=1.50\,,\quad c=7.68\,, \label{eq:paramsanalytic}
\end{align}%
from the fit to Fig.~4 in \cite{Mitter:2014wpa}. 

Furthermore we discuss the gluon self energy in the presence of
magnetic fields at finite temperature in this simplified setup. It is
important to notice that we can facilitate our calculations by the
following argument. The function on the right hand side of
\eq{eq:DSEzeros:1} is a continuous real function of $M_{\rm IR}$ and
approaches $+1$ as $M_{\rm IR}\rightarrow\infty$. Hence it is
sufficient to check whether the expression is negative for $M_{\rm
  IR}=0$, because then it had to pass through zero at some point,
which means that a solution exists.

The gluon self energy receives two important contributions. The first is
through the appearance of fermion loops, which are also present in an
abelian calculation. The fermionic self energy part in lowest Landau
level approximation with $M_{\rm IR}=0$ factorizes 
\begin{align}
  \Pi_f^{\mu\nu}(p)=\alpha eB \exp{\left(-p_\perp^2 /2eB
    \right)}\Pi^{\mu\nu}(p_\parallel,T)\,.\label{eq:gluon-selfEnergy}
\end{align}
Contracting with $P^{\mu\nu}$ in the Landau gauge, we can write the
second term as
\begin{align}
  & \Pi_f(p_\parallel,T)=-8\pi^2\left[ 3-2(1-p_\parallel^2/p^2)
  \right]\nonumber
  \frac{1}{\tau^2}\\[2ex]
  &\hspace{.3cm}\times\int\limits_0^1\dm
  x\sumint\limits_{\tilde{q}_\parallel}\frac{x(x-1)}{\left(
      \tilde{q}_3^2+(2\pi)^2(n+1/2)^2+x(1-x)/\tau^2 \right)^2}\,,
\end{align} 
where
we defined $\tau^2\equiv T^2/p_\parallel^2$. The function can be
evaluated numerically and is very well described by the simple
function
\begin{align}
  \Pi_f(p_\parallel,T)=(1/2\pi) \left[ 3-2(1-p_\parallel^2/p^2)  
  \right] \frac{1}{1 + (4\pi^2/3) \tau ^2}\,.\label{eq:fitSelfenergy}
\end{align}
\Eq{eq:gluon-selfEnergy} and \Eq{eq:fitSelfenergy} state that the
relevant contributions to the self energy stem from $p_\perp^2 \approx
2eB$ and $p_\parallel^2\approx T^2$. Similar as before, the influence
of the magnetic field onto the Yang-Mills sector is not easily
accounted for.  Here we focus on the abelian-like part of the gluon
self energy. As we have investigated before numerically, this is
qualitatively correct and we will use \Eq{eq:scales} to account for
the correct scales.
\begin{figure}[t]
\includegraphics[width=.95\columnwidth]{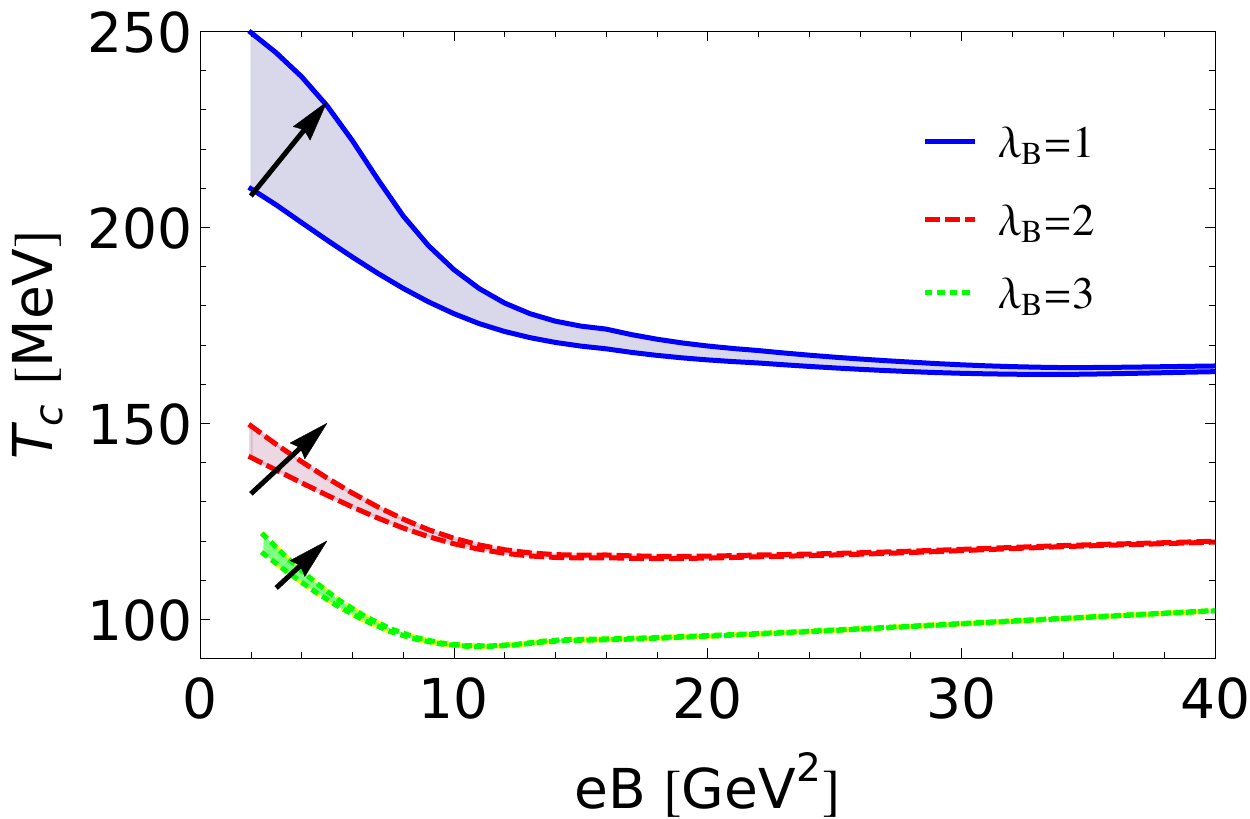}
\caption{Analytic calculation of the critical temperature for the
  chiral phase transition. The bands indicated correspond to $\lambda_T=1$ and
  $\lambda_T=0$. Arrows indicate the direction from $\lambda_T=1$ to
  $\lambda_T=0$.
  }
\label{fig:analyticTc}
\end{figure}
It is well known from Dyson-Schwinger studies \cite{Alkofer:2008tt},
that approximations similar to this semi-bare vertex ansatz
underestimate the strength of chiral symmetry breaking, due to the
negligience of important tensor structures in the vertex, especially
those structures that break chiral symmetry explicitly
\cite{Mitter:2014wpa}. In order to compensate the overall weakness of
the interaction, we allow for a phenomenological parameter $\kappa$ in
front of the integral in \Eq{eq:DSEzeros:1}.

\begin{figure}[t]
  \includegraphics[width=.95\columnwidth]{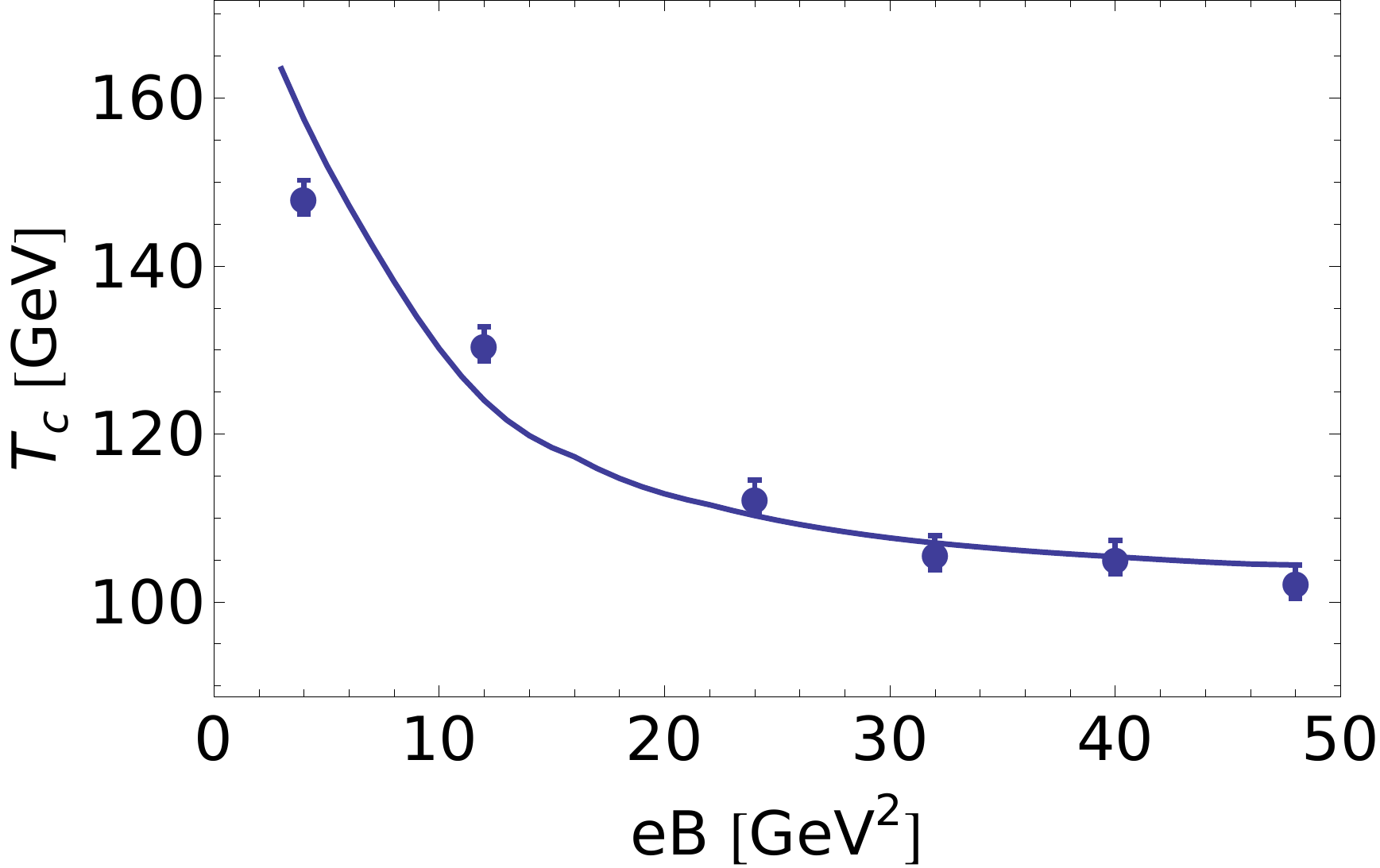}
\caption{Comparison of the critical temperature obtained with our full
  numerical procedure to the simple analytic estimate for
  $\lambda_B=1.1$, $\lambda_T=1$ and $\kappa=1.19$.}
\label{fig:comparisonDSE}
\end{figure}
Using our simple ansatz we can investigate chiral symmetry. In
\Fig{fig:analyticTc} a family of solutions to \Eq{eq:DSEzeros:1} is
shown for various values of $\lambda_B$ and $\lambda_T$, using the
ansatz described above with $\kappa=1.2$ for the two upper curves and
$\kappa=1.4$ for the lower curves.  The choice of $\kappa$ is for
better visualisation only, as the curves can be shifted up and down
using this parameter. 

The observed behavior agrees with that in our numerical study.  It can
be seen from \Fig{fig:analyticTc}, that for small $eB$ inverse
magnetic catalysis is present, while at large $eB$ the the critical temperature 
rises again with the magnetic field, with 
\begin{align}\label{eq:TclargeBDSE}
T_c(B/\Lambda_{\rm QCD}^2\to \infty)\propto \sqrt{ e\,B}\,, 
\end{align}
as one would anticipate from dimensional considerations. This behavior
is universal for all $\lambda_B$ and $\lambda_T$.  We see that the
choice of $\lambda_B$ effects the position of the turning point of the
chiral phase boundary.

With the present analytical considerations the numerical results in
\Fig{fig:numericTc_1} and \Fig{fig:numericTc_2} are readily explained:
they roughly correspond to $\lambda_B\approx 1$, which explains the
relatively large value of $eB$ at the turning point. We see that
already small changes in $\lambda_B$ have a huge effect on this
quantity, see \Fig{fig:analyticTc}. 

In \Fig{fig:comparisonDSE} we have plotted the analytic result with
$\lambda_B=1.1$, $\lambda_T=1$ and $\kappa=1.19$, which agrees well
with the numerical results from scenario 2. Based on the present work
we estimate that $\lambda_B\approx 2-3$ is a realistic choice for the
$B$-dependence of the running coupling, as in our numerical study
quark and gluon propagator turn into their corresponding
$B=0$-propagators at this momentum scale. 

The present analysis reveals the following mechanism: The gauge sector
acquires a $B$-dependence through the feedback of the fermionic
sector. This dependence is responsible for the phenomena called
inverse magnetic catalysis, as has been also observed recently in a
FRG-study within QCD, \cite{Braun:2014fua}. This also explains why it
cannot be seen in model calculation without explicit QCD input. From
\Eq{eq:DSEzeros:1}, \Eq{eq:alpha1} and \Eq{eq:fitSelfenergy} we see
that the gluon screening and the running of the strong coupling (both
by thermal and magnetic effects) are competing with the generic
fermionic enhancement of chiral symmetry breaking in a dimensionally
reduced system. We see from \Fig{fig:analyticTc} that at small
magnetic field screening effects dominate the behavior of the
fermionic self energy, while at asymptotically large fields, thermal
fluctuations are negligible and hence $eB$, as the dominating scale,
drives the phase transition towards higher $T_c$ (magnetic catalysis).

%
%
%
%
\subsection{Four-fermi coupling}\label{sec:QCD-NJL}%
\begin{figure}[t]
\includegraphics[width=.8\columnwidth]{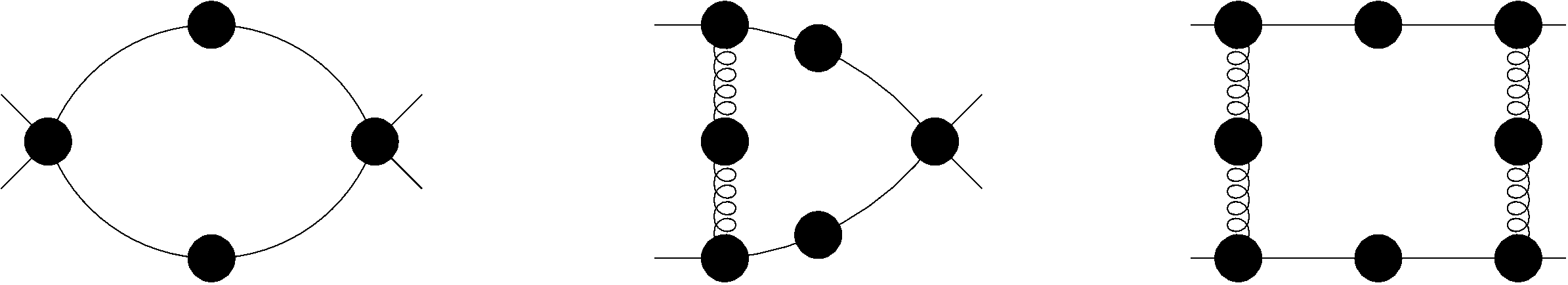}
\caption{Diagrams contributing to the renormalisation group
flow of the four-fermi coupling.}
\label{fig:fourfermiflow}
\end{figure}
For a further analytical grip we also resort to a low energy effective
theory point of view: integrating-out the gapped gluons leads to an
effective four-fermi theory, that is initialized at about the
decoupling scale of the glue sector of $\Lambda\approx 1$
GeV. Previously there have been phenomenological approaches in low
energy effective models to include QCD dynamics as the source of the
inverse magnetic catalysis effect
\cite{Fraga:2013ova,Ferrer:2014qka,Ferreira:2014kpa}. From the point
of view of the FRG for QCD this can be seen as follows
\cite{Braun:2006jd,Braun:2005uj,Braun:2009gm,Braun:2011pp%
  ,Pawlowski:2014aha,Mitter:2014wpa,Braun:2014ata}: At a large
momentum scale $k$ QCD is perturbative, and the 1PI effective action
$\Gamma_k$ in \eq{eq:1PI} is well-described perturbatively. A
four-fermi coupling is generated from the one-loop diagrams (in full
propagators and vertices) encoded in \eq{eq:1PI}, the related diagrams
are depicted in \Fig{fig:fourfermiflow}. In the present discussion we
have dropped diagrams that depend on the $q\bar q-AA$ vertex, $q q\bar
q \bar q-AA$-vertex and the $qqq \bar q\bar q\bar
q$-vertex. Furthermore we assume a classical tensor structure for the
$\bar q A q$-vertex with a coupling $\sqrt{4 \pi \alpha_{s,k}}$, and
only consider the scalar--pseudo-scalar four-fermi vertex
\begin{align}\label{eq:four-fermi}
  \Gamma_{\text{\rm four-fermi}}[q,\bar q,B] &= \frac{1}{2}\, \bar q _i^{a\alpha}
    q_j^{b\alpha}\; \Gamma_{k,ijlm}^{abcd}\;
    \bar q_l^{c\beta} q_m^{d\beta}\,,
\end{align} 
with the scalar--pseudo-scalar tensor structure 
\begin{align}
 \Gamma_{k, ijlm}^{abcd}& = \lambda_k\bigl[
   \delta_{ij}\delta_{lm}\delta^{ab}\delta^{cd}
  + (\rmi\gamma_5)_{ij}(\rmi\gamma_5)_{lm}
  (\tau^n)^{ab}(\tau^n)^{cd} \bigr] \,.
\label{eq:ps-tensor}
\end{align}
The four-fermi term in \eq{eq:four-fermi} can be viewed as the
interaction term of a NJL-type model. Within the approximation to QCD
outlined above the flow of the four-fermi coupling, $\partial_t
\lambda_k$, has the form
\begin{align}\nonumber 
  \partial_t \lambda_k = &- k^2 \lambda_k^2 F_{\lambda}(G_q) - 
  \lambda_k\alpha_{s,k} F_{\lambda\alpha_s}(G_q,G_A) \\[2ex]�
& - \0{\alpha_{s,k}^2}{k^2}
  F_{\alpha_s^2}(G_q,G_A)\,, 
\label{eq:dtlambda}\end{align} 
with positive coefficients
$F_{\lambda},F_{\lambda\alpha_s},F_{\alpha_s}$. The respective
diagrams are depicted in \Fig{fig:fourfermiflow}. The different
classes of diagrams in \Fig{fig:fourfermiflow} depend on combinations
of gluon and quark propagators, $G_A$ and $G_q$ respectively. 

\begin{figure}[t]
\includegraphics[width=.85\columnwidth]{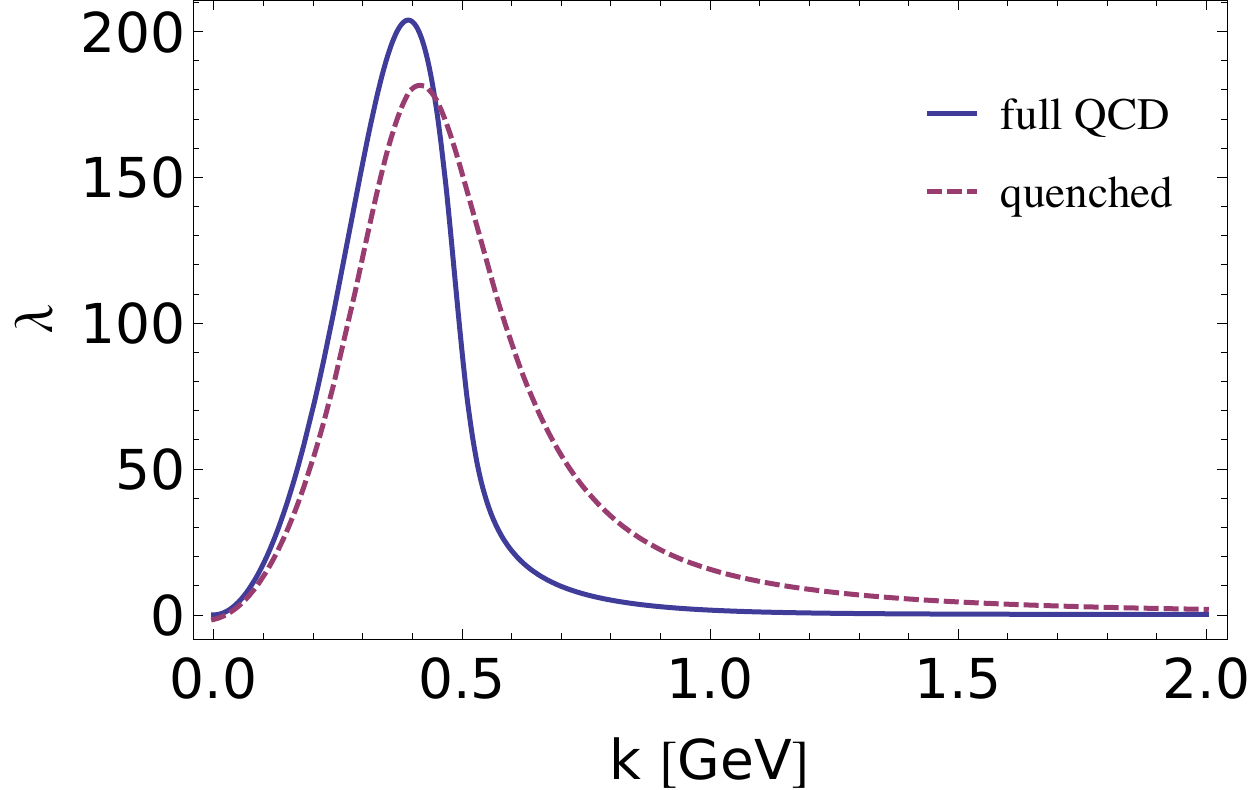}
\caption{Scalar--pseudo-scalar four-fermi coupling in the vacuum, $T=0$, $B=0$, computed with 
  quantitatively reliable QCD-flows in quenched QCD,
  \cite{Mitter:2014wpa}, and with qualitative full QCD flows, \cite{Braun:2014ata}.} 
\label{fig:comp_lambda_quenched-vs-full}
\end{figure}

The four-fermi coupling $\lambda_k$ in two-flavor QCD at $T=0$ has
been quantitatively computed (including its momentum-dependence) in
quenched QCD with the FRG in \cite{Mitter:2014wpa}, and in a more
qualitative approximation (without its momentum-dependence) in fully
dynamical QCD in \cite{Braun:2014ata}. The respective results are
depicted in \Fig{fig:comp_lambda_quenched-vs-full}. As expected, the
couplings have a similar dependence and maximal strength. However, the
slope of the coupling in the qualitative computation in the peak
regime relevant for chiral symmetry breaking is bigger for the
qualitative computation. This can be traced back to the missing
momentum-dependencies, whose lack artificially increases the locality in
momentum space and in the cutoff scale. Hence, guided by the experience gained in the
DSE-computations we expect the slope to play a large r$\hat{\rm o}$le
and we shall use the quantitative quenched results for $\lambda_k$ and
$\alpha_s$ in our present computations. We shall further comment on
the differences in the next Section.

For large cutoff scales $k$ the propagators approach the classical
propagators. The current quark mass at these scales is negligible and
only the cutoff scale is present, if temperature and magnetic field
are considered small relative to the cutoff scale. Then the
dimensionless $Fs$ are simple combinatorial factors. For optimized
regulators, \cite{Litim:2000ci}, they are given as
\begin{align} 
  F_{\lambda}=4N_c\,,\quad
  F_{\lambda\alpha_s}=12\frac{N_c^2-1}{2N_c}\,, \quad 
  F_{\alpha_s^2}=\frac{3}{16} \frac{9N_c^2-24}{N_c}\,,
\label{eq:Fasympt}\end{align}
in the vacuum, see
e.g.\ \cite{Braun:2011pp,Mitter:2014wpa,Braun:2014ata} for more
details.  For small enough cutoff scales $k$ the gluonic diagrams
decouple due to the QCD mass gap. In the Landau gauge this can be
directly seen with the gapping of the gluon propagator. For $T=0,\,B=0$ this entails
\begin{align}
\label{eq:gluegap}
p^2 G_A(p^2\lesssim \Lambda^2) \propto p^2/m_{\rm gap}^2\,.
\end{align} 
with $\Lambda\approx 1$ GeV. We emphasize that \eq{eq:gluegap} only
reflects the mass gap present in the Landau gauge gluon propagator,
the gluon propagator is not that of a massive particle, see e.g.\
\cite{Fischer:2008uz}. For momentum scales $p^2\lesssim \Lambda^2$
this approximately leaves us with an NJL-type model with the action
\begin{align}\label{eq:Gamma_NJL}
  \Gamma_{\rm NJL}[q,\bar q,B] &= \int_x  \bar
    q\,\rmi\feyn{\partial} q + \Gamma_{\text{\rm four-fermi}}[q,\bar q,B] \,,
\end{align} 
with the scalar--pseudo-scalar four-fermi interaction defined in
\eq{eq:four-fermi}.  In the presence of a magnetic field this model
including fermionic fluctuations has been investigated in
\cite{Fukushima:2012xw} within the FRG. Here we shall use the
respective results within the lowest Landau level approximation. Then
$T_c$ shows an exponential dependence on the dimensionful parameter
$eB$
\begin{align}
  T_c=0.42\Lambda\exp{\left( -\frac{2\pi^2}{N_c\lambda_\Lambda
\sum\limits_f|q_feB|}\right)}\label{eq:NJLjanoriginal}.
\end{align} 
The well-known exponential dependence of $T_c$ on the four-fermi coupling
$\lambda_\Lambda$ already explains the large sensitivity of the scales
of magnetic calatysis and inverse magnetic catalysis to details of the
computation. \Eq{eq:NJLjanoriginal} is valid for large magnetic field
and for $\Lambda\ll m_{\rm gap}^2$, that is deep in the decoupling
regime of the gluons. An estimate that also interpolates to small
magnetic fields is given by
\begin{align}
  T_c=0.42\Lambda \exp {  \left(-\0{1}{ \Co_{\Lambda} \lambda_\Lambda}\right)} \,,
\label{eq:NJLestimate} 
\end{align}
with 
\begin{align} 
  \Co_{k}(B)= \0{N_c}{2 \pi^2}\left( \sum_f| q_f e\,B| + c_1\, k^2 \right)\,,\quad
  {\rm with}\quad c_1= 3\,,
\label{eq:ck} 
\end{align} 
where $c_1$ has been adjusted to reproduce $T_c(B=0)\approx 158$ MeV. While \Eq{eq:ck}
resembles a lowest Landau level approximation, it is actually an expansion in $B$.
Using this ansatz we can describe the behavior of the phase transition
on scales below $1\text{ GeV}^2$ qualitatively, while the $B=0$ limit is fixed.

It is also well-known that for $k\gg m_{\rm gap}$ the flow of the four-fermi
coupling is driven by the gluonic diagrams summed-up in
$F_{\alpha_s}$: for large scales we can set $\lambda_{k\gg m_{\rm
    gap}}\approx 0$. The gauge coupling is small, $\alpha_{s,{k\gg
    m_{\rm gap}}}\ll1$ and the flow gives $\lambda_k\propto
\alpha_s^2$. This entails that the diagrams with four-fermi couplings
are suppressed by additional powers of $\alpha_s$, and the four-fermi
coupling obeys
\begin{align}\label{eq:QCDwomatterflow} 
  \partial_t\lambda_{ \text{\tiny glue} ,k} = -\0{\alpha_{s,k}^2}{k^2}
  F_{\alpha_s}(G_q,G_A)\,,
\end{align}
where the subscript 'glue' indicates that the flow is driven by glue
fluctuations. As discussed before, for $k\gg m_{\rm gap}$ we have
classical dispersions for quark and gluon, and the diagrammatic factor
$F_{\alpha_s}$ is a constant, see \eq{eq:Fasympt}. The strong coupling
$\alpha_{s,k}$ has the form \eq{eq:alpha1} with $z\propto
k$. Integrating \eq{eq:QCDwomatterflow} with \eq{eq:alpha1} gives
\begin{align}\label{eq:laglue}
  \lambda_{ \text{\tiny glue} ,k}\propto \0{\alpha_{s,k}^2}{2 k^2}
  F_{\alpha_s}(G_q,G_A)\,.
\end{align}
where an estimate for the $B$-dependence of the gluonic diagram in
$F_{\alpha_s}$ is given in Appendix~\ref{app:Falpha}.  

\begin{figure}[t]
\includegraphics[width=.9\columnwidth]{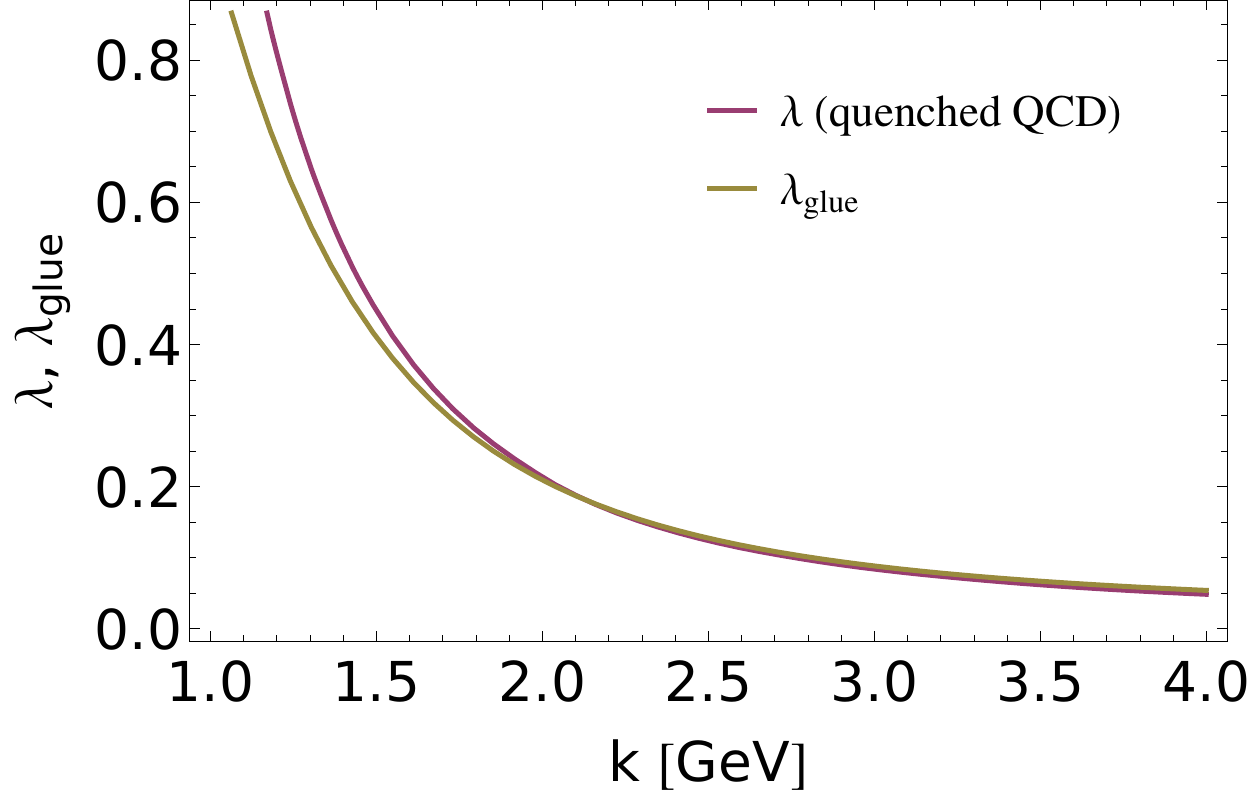}
\caption{Scalar--pseudo-scalar four-fermi coupling at $T=0,\, B=0$
  computed with quantitatively reliable QCD-flows in quenched QCD,
  \cite{Mitter:2014wpa}, in comparison to $\lambda_{\text{\tiny glue}}
  $ computed from \eq{eq:laglue}. }
\label{fig:fourfermiQCDglue}
\end{figure}

\begin{figure}[t]
\includegraphics[width=.95\columnwidth]{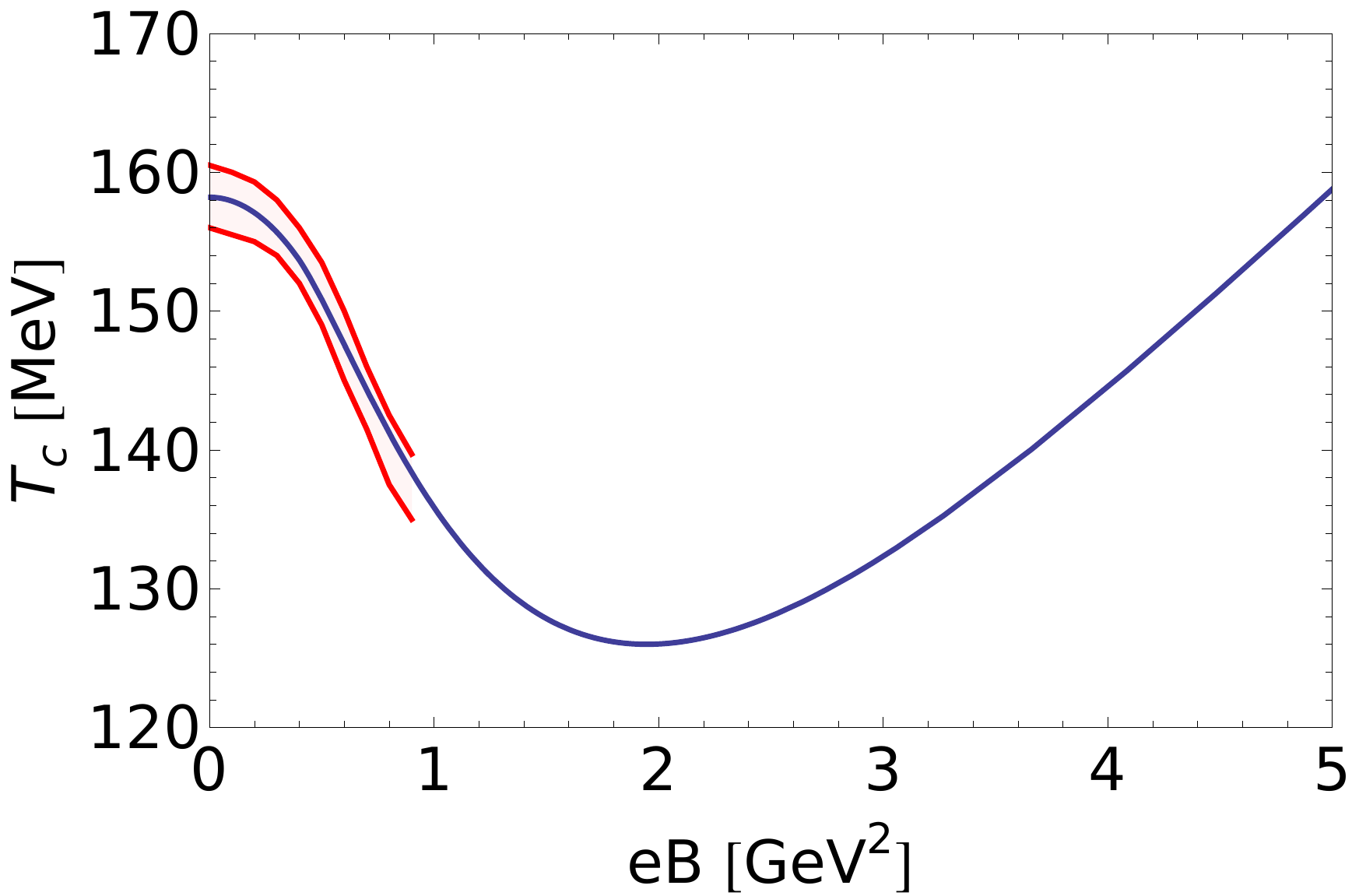}
\caption{Comparison of the chiral transition temperature obtained
  within the simple mean field NJL estimate \Eq{eq:tcsimple} to the lattice
  results of \cite{Bali:2011qj} (see their Fig.~10).}
\label{eq:tcsimple}
\end{figure}

At vanishing magnetic field $\lambda_{ \text{\tiny glue} ,k}$ agrees
well with the full result for the four-fermi coupling in
\cite{Mitter:2014wpa} for $k\gtrsim 2$ GeV, see
\Fig{fig:fourfermiQCDglue}. Below $k\approx 2$ GeV, $\lambda_{
  \text{\tiny glue} ,k}$ is increasingly smaller than the full
scalar--pseudo-scalar four-fermi coupling in quenched QCD.
In this intermediate range, where all diagrams contribute, we write the
resulting coupling within a resummed form that captures already the
fermionic diagram proportional to $F_\lambda$,
\begin{align}\label{eq:lambdaresum}
  \lambda_k= \0{\bar \lambda_k }{1 - \bar \Co_{k}\bar \lambda_k
  }\,,\quad {\rm with} \quad \bar \Co_{k} =  \int_k^\Lambda dk'\,k'
  F_\lambda(G_q)\,.
\end{align} 
The resummed form in \eq{eq:lambdaresum} already reflects the matter
part of the flow in \eq{eq:dtlambda} which is the term proportional to
$\partial_t {\lambda_k}$. The other terms add up to 
\begin{align}\label{eq:dtbarlambda}
  \partial_t \bar\lambda_k= - (1 - \bar \Co_{k}\bar \lambda_k)^2
  \left( \lambda_k\alpha_{s,k} F_{\lambda\alpha_s}+
    \0{\alpha_{s,k}^2}{k^2}\right)\,.
\end{align}
For $\bar \Co_{k}\bar \lambda_k\ll 1$ the flow of
$\bar\lambda_k$ boils down to \eq{eq:QCDwomatterflow}. For $\bar
\Co_{k}\bar \lambda\to 1$ the flow in \eq{eq:dtbarlambda} tends
towards zero. In this regime the four-fermi coupling grows large and
the matter flow dominates. Hence, for the present qualitative analysis
we simply identify $\bar\lambda$ with the glue $\lambda_{\text{\tiny
    glue}}$, \eq{eq:laglue}, up to a prefactor,
\begin{align}\label{eq:barla}
\bar \lambda_{k}= Z_{\lambda} \lambda_{ \text{\tiny glue} ,k}  \,.
\end{align}
The prefactor $Z_\lambda$ accounts for the fact that we have used
results of quantitative QCD-flows \cite{Mitter:2014wpa} for the strong
coupling which also includes wave function renormalisations for the
quarks. In the current model considerations without wave function
renormalisation and further simplifications this has to be accounted
for. For the same reason the normalisation $0.42\, \Lambda$ related to a
four-fermi flow with an optimised regulator has to be
generalised. Moreover, the prefactor $\bar c_{\lambda,k}$ is the
integrated four-fermi flow already present in \eq{eq:NJLestimate} up
to an overall normalisation accounting for the model
simplifications. We choose 
\begin{align}\label{eq:c2c3} 
\bar \Co_{k}(B) =c_3\, \Co_{k}(B)\,,\quad {\rm and}\quad 
0.42 \,\Lambda \to 0.42\, \Lambda\, \exp\left(c_2-c_3\right)\,, 
\end{align}
and arrive at
\begin{align}
  T_c=0.42\Lambda \exp\left( -\0{1}{\Co_{\Lambda} \bar
      \lambda_\Lambda}+c_2\right)\,, 
 \label{eq:NJLfull} 
\end{align} 
with $\Co_{\Lambda}$ as given in \eq{eq:ck} and $\bar\lambda$ in
\eq{eq:barla} and \eq{eq:laglue}. Note that the parameter $c_3$ has
dropped out. Its value can be adjusted to achieve a quantitative
agreement of $\eq{eq:laglue}$ with the QCD result in
\cite{Mitter:2014wpa} with
\begin{align}\label{eq:valuec3} 
c_3 =  \0{1}{2 Z_\lambda}\,, 
\end{align}
where the factor $1/Z_\lambda$ simply removes the mapping factor
adjusting for the missing wave function renormalisations in the model
computation. This quantitative agreement strongly supports the
reliability of the approximate solution to the flow equation given by
\eq{eq:lambdaresum} in the intermediate momentum regime that is of
importance for the current considerations. The remaining parameters
are fixed as follows,
\begin{align}\label{eq:parameters}
Z_\lambda=2.2\,, \quad  c_1=3\,, \quad c_2=1.4\,.
\end{align}
The parameter $c_1$ has already been adjusted to meet $T_c(B=0)\approx
158$ MeV, see \eq{eq:NJLestimate} and \eq{eq:ck}. The parameter $c_2$
re-adjusts the overall scale $0.42\, \Lambda \to 0.42\, \Lambda \exp
c_2 = 1.7\Lambda$. As already discussed above, it depends on the
regulator and the approximation at hand. It reflects the dependence on
the renormalisation group scheme. Similarly to $c_1$ it is fixed with
$T_c(B=0)\approx 158$ MeV, and is a function of the overall
normalisation of the four-fermi coupling $Z_\lambda$.  The latter is
the only free parameter left. In \eq{eq:parameters} we use the value
that reproduces the lattice results, see \Fig{eq:tcsimple}. We
emphasise that no other parameter is present that allows to shift the
minimum in $T_c$, the latter being a prediction. 

Obviously, the effect seen in our numerical and analytic DSE-study, is
also present in the analytic approach to the dynamics of the
four-fermi coupling, including a direct grip on the underlying
mechanisms. We see that the non-monotonous behavior, i.e. the delayed
magnetic catalysis, \cite{Braun:2014fua,Ilgenfritz:2013ara}, is
already present at smaller scales compared to \Fig{fig:numericTc_1}
and \Fig{fig:numericTc_2}, while the lattice results are reproduced.

In turn, for asymptotically large magnetic field, the critical temperature 
runs logarithmically with $B$, 
\begin{align}\label{eq:TclargeBFRG}
T_c(B/\Lambda_{\rm QCD}^2\to\infty) \propto \ln  B/\Lambda_{\rm QCD}\,, 
\end{align}
related to a double-log--dependence on $B$ of the exponent. Due to the
qualitative nature of the approximation of the $B$-dependence of the
gluon propagator it cannot be trusted for asymptotically large
$B$. Indeed, \eq{eq:TclargeBFRG} has to be compared to
\eq{eq:TclargeBDSE} within the analytic DSE-approach predicting a
square root dependence. Note that in the latter computation the quark
vaccum polarisation is included selfconsistently at large $B$ even
though the backreaction on the pure glue loops in \Fig{fig:gluonDSE}
is neglected. Still this indicates the validity of the square root
dependence, even though a definite answer to this question requires
more work.

\subsection{Discussion of scales \& mechanisms}\label{sec:mech}
With the findings of the last two sections we have achieved an
analytic understanding of the mechanisms at work. The decrease of
$T_c$ for small magnetic fields, the increase of $T_c$ for larger
fields, as well as the related magnetic field regimes can now be
understood.  In particular this concerns the magnetic field
$B_{\text{\tiny min}}$, where $T_{c}(B_{\text{\tiny min}})$ is at its
minimum.  This is the turning point between increasing and decreasing
$T_c(B)$.

Magnetic catalysis relates to the dimensional reduction
due to the magnetic field in diagrams with quark correlation functions
leading to an increase of the condensate. At finite temperature the
catalysis due to the dimensional reduction is accompanied by a thermal
gapping of the quarks that counteracts against the magnetic catalysis
effects. In total this leads to a rise of both, the chiral condensate
and the critical temperature, if the magnetic field dependence of the
involved couplings is sufficiently small. As the magnetic field also
sets a momentum scale of the physics involved, this scenario holds
true for sufficiently large magnetic field strength $e
B/\Lambda_{\text{\tiny QCD}}^2\gg 1$, where the $B$-dependence of the
couplings can be computed (semi-)-perturbatively. This explains the 
regime of delayed magnetic catalysis. 

The above discussion of the standard scenario already entails that
rapidly changing couplings are required for a decreasing $T_c$. The
couplings involved are the scalar--pseudo-scalar four-fermi coupling
$\lambda_k$ and the strong coupling $\alpha_{s,k}$, where $k$ sets the
momentum scale. Both are rising rapidly towards the infrared for
momentum scales $k\lesssim 4-10$ GeV, for $\lambda_k$ see
\Fig{fig:fourfermiQCDglue}. In this regime chiral symmetry breaking
and confinement is triggered and takes place in QCD at vanishing
magnetic field. Switching on the magnetic field increases the relevant
momentum scale $k^2 \propto e\, B$ and hence decreases $\lambda$ and
$\alpha_s$. The condensate still grows with $B$ as the $B$-enhancement
in the broken phase is still present, only $T_c$ decreases.

Our results from the analytic approach to the quark gap equation,
presented in \Fig{fig:analyticTc}, support these findings.  The
position of the turning point $B_{\text{\tiny min}} $ in both the full
numerical as well as the analytic analysis of the gap equation depends
crucially on the magnetic field and temperature dependence of the
quark gluon vertex, see \Fig{fig:analyticTc}.  When contrasted with
the quantitative FRG results of $\alpha_s$ in \cite{Mitter:2014wpa},
the strong coupling in \eq{eq:DSE:vtx} decays considerably slower
towards the UV. In turn, the couplings in the qualitative FRG study
for full QCD, \cite{Braun:2014ata} have a steeper decay, for the
four-fermi coupling see \Fig{fig:fourfermiQCDglue}. Seemingly, this
already explains the large value of $B_{\text{\tiny min}} $ in the
current DSE-study as well as the small value of $B_{\text{\tiny min}}
$ in \cite{Braun:2014fua}, which uses approximations similar to
\cite{Braun:2014ata}. Note however, that we have used the quenched
quantitative $\alpha_s$ in the analytic DSE-study which agrees well
with the numerical DSE result for $\lambda_B\approx 1$. 

In summary we have identified the physics mechanisms behind the $T-B$
phase diagram from our full QCD calculations. Moreover,
\Fig{eq:tcsimple} suggests a turning point for $e\,B_{\text{\tiny
    min}} \approx 1.5 - 10$ GeV${}^2$, the large regime for
$e\,B_{\text{\tiny min}} $ being related to the exponential dependence
on the couplings. Evidently, the effects observed depend on a
sensitive balance of different scales and parameters. Hence, further
studies are required to fully uncover the intricate underlying
dynamics. Very recent findings in AdS/QCD models, \cite{Mamo:2015dea},
indicate an inverse magnetic catalysis behavior up to $eB\approx 4
\text{ GeV}^2$, which supports our findings.

\section{Conclusions}%
We have investigated the chiral phase structure of QCD at finite
temperature in the presence of an external magnetic field. Our study
resolves the discrepancy between recent lattice and continuum
calculations at magnetic fields below $1 \text{ GeV}^2$, see also
\cite{Braun:2014fua}. We confirm the inverse magnetic catalysis effect
seen in lattice studies at small $B$. At larger $B$ we see that
magnetic catalysis is restored, with $T_c\propto
\sqrt{eB}$. Indications for the turnover behavior have already been
found in \cite{Braun:2014fua}, and in \cite{Ilgenfritz:2013ara} within
two-color lattice-QCD. %
We hope that further lattice calculations in full QCD at the scales
discussed here will become feasible soon.

The reason for this non-monotonous behavior are screening effects of
the gauge sector, i.e. modifications of the gluon self energy, as well
as the strong coupling $\alpha_s$ in the presence of magnetic fields.
Moreover we have investigated the nature of the chiral transition at
finite magnetic field. 

Apart from the $B$-dependence of the critical temperature, we observe
that the phase transition in the chiral limit turns smoothly into a
crossover with rising $B$. Notably, we find a non-degeneracy in the
phase transition which is due to the explicit isospin breaking caused
by the different electric charges of up and down quark. This
non-degeneracy might lead to phenomenological consequences in
experimental studies of the QCD phase diagram with non-central
heavy-ion collisions, as there might be a mixed phase 
between the up and down quark transitions. Recent lattice calculations 
\cite{Endrodi:2015oba} support the possibility of a non-degenerate 
chiral phase transition. 

In addition, our calculations show that, due to this isospin breaking,
there is a step-like behavior in the up quark condensate triggered by
the chiral transition of the down quark. While this is an significant
effect in the chiral limit it smoothens out rapidly with increasing
current quark mass. Physical current quark masses are in the
transition regime, and this effect might have phenomenological
consequences. To our knowledge, this is a novel effect in the QCD
phase diagram and it certainly deserves further investigation.

We have used analytic studies of the quark gap equation and the
dynamics of the four-fermi coupling for an investigation of the
physics mechanisms behind (inverse) magnetic catalysis. The results
are discussed at length in the previous Section~\ref{sec:mech},
leading to a rough prediction of the turning point at
$e\,B_{\text{\tiny min}} \approx 1.5 - 10$ GeV. Our investigations
highlight the rich phenomenology of QCD matter in external magnetic
fields, which motivates further studies, e.g.\ at finite chemical
potential, towards more realistic descriptions of
matter under extreme conditions. Recent studies \cite{Nishiyama:2015fba} have suggested
even richer QCD phase structures in the presence of magnetic fields.

\acknowledgments 
We thank J.~Braun, C.S.~Fischer, K.~Fukushima, W.A.~Mian,
M. Mitter, S.~Rechenberger, F.~Rennecke and N.~Strodthoff for
discussions and work on related subjects. This work is supported by
the Helmholtz Alliance HA216/EMMI and the grant ERC-AdG-290623. NM
acknowledges support by the Studienstiftung des Deutschen Volkes. %
%
%
%
%
\begin{appendix}
  \section{Gluon Propagator and Quark Gluon vertex from Dyson
    Schwinger studies\label{app:propsandvertex}}%
Here we discuss the truncation scheme for the quark gap equation and
the gluon propagator, based on
\cite{Fischer:2010fx,Fischer:2012vc}. The quark gluon vertex is taken
as $\Gamma^{\mu}=z_{qgq}\gamma^{\mu}$, with
\begin{align}
  z_{qgq}(Q^2)=&\frac{d_1}{d_2+Q^2}\\[2ex] 
& +\frac{Q^2}{\Lambda^2+Q^2}\left(
    \frac{\beta_0\alpha(\mu) \log{Q^2/\Lambda^2+1}}{4\pi}
  \right)^{2\delta}\,,\label{eq:DSE:vtx}
\end{align}
containing the parameters
\begin{align}
  d_1=7.9\text{ GeV}^2\, & \qquad d_2=0.5\text{ GeV}^2\,,\nonumber \\[2ex] 
  \delta = -18/88\,, & \qquad\Lambda =  1.4\text{ GeV}\,.
\end{align}
Here the scales must be identified correctly in order to capture the
correct dependence with $T$ and $eB$. We take $Q$ to be the symmetric
momentum $Q^2=(q^2+p^2+(q-p)^2)/3$ at the vertex with
$Q^2=Q_3^2+Q_0^2+Q_\perp^2$, where $Q_0^2=(2\pi T)^2$ if $Q_0^2<(2\pi
T)^2$ and $Q_\perp^2=2|eB|$ if $Q_\perp^2<2|eB|$. We note that this
roughly corresponds to an identification of scales as in section
\ref{sec:DSEan} with $\lambda_B\approx 1$, although the present vertex
is clearly more sophisticated as it includes momentum dependencies and
thereby generic $eB$ effects.  For a current overview of the quark
gluon vertex in Dyson-Schwinger truncations see
\cite{Alkofer:2014taa,Williams:2014iea}.  Furthermore in order to be
able to solve the gluon Dyson-Schwinger equation we rely on lattice
input for the Yang-Mills part, which we then "dress" with magnetic
field effects, as described above.  The reliability of this truncation
was already discussed in detail at finite temperature
\cite{Fischer:2012vc} and utilized in the presence of magnetic fields
before \cite{Mueller:2014tea}. The lattice fit is given by
\begin{align}
  Z_\text{YM}^{-1}(Q^2)=\frac{Q^2\Lambda^2}{(Q^2+\Lambda^2)^2}\Big[
  \left( \frac{c}{Q^2 +a\Lambda^2}\right)^b\nonumber \\[2ex] 
+  \frac{Q^2}{\Lambda^2}\left( \frac{\beta_0
      \alpha(\mu)\log{Q^2/\Lambda^2+1}}{4\pi}\right)^\gamma
  \Big]\,,
\label{eq:gluonQuenched}
\end{align}
with 
\begin{align}
  \Lambda = 1.4\text{ GeV}\,, &\quad c = 11.5\text{ GeV}^2\,,
  \nonumber\\[2ex]
  \beta_0 = 11N_c/3\,, &\quad \gamma = - 13/22\,,
\end{align}
where $\alpha(\mu)=0.3$ and $a$ and $b$ are temperature dependent
parameters, which can be found in \cite{Fischer:2010fx}. As discussed
before the Dyson-Schwinger truncation scheme can be related to the
skeleton expansion done in our analytic estimate, which was motivated
by renormalisation group invariance
\begin{align}
  4\pi\alpha_s(Q^2)r_\text{IR}(Q^2)\frac{P_{\mu\nu}}{Q'^2+\Pi}\equiv
  \frac{P_{\mu\nu}}{Z_\text{YM}Q'^2+\Pi_f}z_{qgq}\,,
  \label{eq:DSEKernel}
\end{align}
where the sum over different polarisation tensor components is
implied. The right hand side actually serves as the input to our
numerical study, while the different components of $\Pi$ are
determined dynamically from solving the gluon Dyson-Schwinger
equation.

\section{Magnetic field dependence of the four-fermi coupling from QCD}
\label{app:Falpha}
As we have discussed in Section~\ref{sec:QCD-NJL} the value of the NJL
coupling $\lambda$ at the intrinsic cutoff scale of the model is
determined by QCD dynamics. At large scales the dynamics of $\lambda$
is driven by the rightmost diagram shown in
\Fig{fig:fourfermiflow}. Within simplifications we will motivate the
functional dependence of this diagram on temperature and the magnetic
field.  In the lowest Landau level approximation the quarks are
constraint to the t-z plane denoted by $(\parallel)$, whereas the
gluons propagate in all four dimensions $(\parallel,\perp)$. We write
the gluon box diagram in \Fig{fig:fourfermiflow} at zero external
momentum as
\begin{widetext}
\begin{align}
F_{\alpha_s} (e\,B\geq 0.3\, {\rm GeV})
 \simeq 4.5\, e\,B \int_0^\infty
  d q_\parallel\,,\0{q_\parallel}{q_\parallel^2+
    m_q^2+\alpha_s eB c_q}
  \int_0^\infty d {q_\perp}\,, \0{q_\perp}{[q_\perp^2
    +q_\parallel^2+m_A^2 + eB \alpha_s 
    c_A]^2}\,,  \label{eq:approx_lambda}
\end{align}
\end{widetext}
where $\alpha_s$ is given as \Eq{eq:alpha1}.

For $e\,B< 0.3$ \eq{eq:approx_lambda} is smoothly (quadratic fit)
extrapolated to $e\, B=0$ with minimising the $e\,B$-dependence. The
flavor, color and Dirac tensor indices have been contracted, and the
comparison with the results for $\lambda$ in quenched QCD shown in 
\Fig{fig:fourfermiQCDglue} shows that the prefactor resulting
from the tensor contract is approximately $4.5$.
We have written the propagators in a semi-perturbative form with
medium dependent mass terms. Further we have taken $m_A\approx 1\text{ GeV}$
as the decoupling scale, $m_q\approx 300 \text{ MeV}$ in the chiral
broken phase and $c_A=c_q=1$. Strictly speaking both masses are larger
than 1 GeV as we have to add the cutoff masses $\propto \Lambda^2$. We
have chosen smaller masses in order to also potentially have access to
the infrared domain $k\to 0$, where the constituent quark mass is of
the order $0.3$ GeV and the gluonic mass gap is of the order $1$ GeV.
Furthermore we have approximated the Matsubara sum by an integration,
due to the small level spacing compared to the magnetic field. This
approximation does not hold small $eB$, but \eq{eq:approx_lambda} is
only used for $e\,B\geq 0.3$ GeV. \Eq{eq:approx_lambda}
includes the correct dependence on $\alpha_s$ as well and thus
captures $eB$ and $T$ effects qualitatively. The model parameters in
Section~\ref{sec:QCD-NJL} allow us to reproduce the quantitative
behavior of the chiral transition temperature and a more elaborate
version of \Eq{eq:approx_lambda} does not give much greater insight.
Apart from the agreement with the $T_c$ results from lattice calculation,
 \Fig{fig:fourfermiQCDglue}
shows that quantitatively reliable results from QCD-flows in
quenched QCD \cite{Mitter:2014wpa} are reproduced. 

\end{appendix}

\bibliography{bib_mag.bib}

\end{document}